\documentclass[aps,prb,twocolumn,superscriptaddress,floatfix,showpacs]{revtex4}

\usepackage{latexsym,amsmath,amssymb,amsfonts,mathbbol,graphicx}
\usepackage{dcolumn}
\usepackage{bm}
\usepackage[T1]{fontenc}
\usepackage[latin1]{inputenc}
\usepackage[usenames,dvips]{color}

\begin{document}

\title{Dynamical critical scaling and effective thermalization in
quantum quenches: \\ the role of the initial state}

\author{Shusa Deng} \affiliation{\mbox{Department of Physics and
Astronomy, Dartmouth College, Hanover, New Hampshire 03755, USA}}

\author{Gerardo Ortiz}
\affiliation{\mbox{Department of Physics, University of Indiana,
Bloomington, Indiana 47405, USA}}

\author{Lorenza Viola} \affiliation{\mbox{Department of Physics and
Astronomy, Dartmouth College, Hanover, New Hampshire 03755, USA}}

\date{\today}
\begin{abstract}
We explore the robustness of universal dynamical scaling behavior in a
quantum system near criticality with respect to initialization in a
large class of states with finite energy.  By focusing on a
homogeneous XY quantum spin chain in a transverse field, we
characterize the non-equilibrium response under adiabatic and sudden
quench processes originating from a pure as well as a mixed excited
initial state, and involving either a regular quantum critical or a
multicritical point.  We find that the critical exponents of the
ground-state quantum phase transition can be encoded in the dynamical
scaling exponents despite the finite energy of the initial state.  In
particular, we identify conditions on the initial distribution of
quasi-particle excitation which ensure Kibble-Zurek scaling to
persist.  The emergence of effective thermal equilibrium behavior
following a sudden quench towards criticality is also investigated,
with focus on the long-time expectation value of the quasi-particle
number operator.  Despite the integrability of the XY model, this
observable is found to behave thermally in quenches to a regular
quantum critical point, provided that the system is initially prepared
at sufficiently high temperature.  However, a similar thermalization
behavior fails to occur in quenches towards a multi-critical point.
We argue that the observed lack of thermalization originates in this
case in the asymmetry of the impulse region that is also responsible
for anomalous multicritical dynamical scaling.
\end{abstract}

\maketitle

\section{Introduction}
\label{intro}

Characterizing the non-equilibrium dynamics of quantum many-body
systems is of central significance to both condensed-matter physics
and quantum statistical mechanics.  A quantitative understanding of
non-equilibrium quantum phase transitions (QPTs) is, in particular, a
fundamental prerequisite for uncovering and controlling quantum phases
of matter \cite{Sachdev,Vojta}, as well as for assessing the
complexity of quantum annealing or adiabatic algorithms
\cite{Nishimori,QC}.  Unlike standard phase transitions which are
driven by a change in temperature, QPTs are driven entirely by quantum
fluctuations at zero temperature. They nevertheless share with their
classical counterpart the generic feature of universality: in
equilibrium, the critical properties of a system sufficiently close to
a quantum critical point (QCP) depend only on a few relevant
characteristics such as its symmetry and dimensionality, thus defining
the universality class to which the corresponding (continuous) QPT
belongs. The universality class is distinguished by a small set of
critical exponents -- for instance, $\nu$ and $z$, describing the
power-law divergence of the characteristic length scale and the
vanishing of the characteristic energy scale, respectively
\cite{Sachdev}.  In a non-equilibrium scenario, the system can be
driven across a QCP {\em dynamically}, that is, through an explicit
time-dependence of one or more control parameters in the underlying
many-body Hamiltonian.  This naturally prompts a number of questions:
to what extent can universal quantum scaling laws persist out of
equilibrium and be solely specified in terms of the equilibrium phase
diagram?  Conversely, how does quantum criticality influence the
ability of a system to relax back to equilibrium?

Historically, the first theoretical studies in these directions date
back to the pioneering work by Barouch and coworkers
\cite{Barouch1,Barouch2,Barouch3} which led, in particular, to the
discovery of ``non-ergodic'' behavior in the zero-temperature
long-time magnetization of a driven XY spin chain \cite{Barouch2}.  In
recent years, the demand for quantitatively addressing the above broad
questions has heightened under the impetus of experimental advances in
manipulating ultracold atomic gases, which are enabling the unitary
dynamics of many-body quantum systems to be explored with an
unprecedented level of coherent control and isolation from the
environment \cite{ExpQPT,ExpTherm}.  As a result, non-equilibrium
quantum critical physics is being actively investigated both from a
theoretical and experimental standpoints.

In this framework, an important step forward is provided by the
prediction of universal behavior in {\em adiabatic} dynamics near a
QCP based on the {\em Kibble-Zurek scaling} (KZS) argument
\cite{Zurek} (see also Ref. \onlinecite{Polk1} for related independent
work and Ref. \onlinecite{JacekRev} for a recent review).  Originally
introduced in the context of classical (finite-temperature) phase
transitions in cosmology \cite{Kibble}, the KZ argument rests on the
basic intuition that, irrespective of how slowly a system is driven
across a continuous phase transition, adiabaticity is necessarily lost
in the thermodynamic limit due to the vanishing energy gap at the
critical point. Qualitatively, this determines typical time and length
scales, $\hat{t}$ and $\hat{\xi}$ respectively, that characterize the
adiabatic-to-diabatic crossover and, since ``order'' cannot be
established on distances larger than $\hat{\xi}$, results in the
formation of a domain structure and the generation of a {\em finite}
density of ``topological defects'' in the system.  Quantitatively, let
the time-dependence in the quantum-mechanical Hamiltonian $H(t)$ be
introduced through a control parameter $\lambda (t)$, with
$\lambda_c\equiv \lambda(t_c)$ corresponding to the crossing of an
isolated QCP at time $t_c$, which can be taken to be $t_c\equiv 0$
without loss of generality.  If the system is initially ($t=t_0$) in
the ground state, its ability to adiabatically adjust to $H(t)$ is
determined by the condition that the typical time scale $\tau(t)\equiv
|(\lambda(t)-\lambda_c)/\dot{\lambda}(t)|$ associated with the applied
control be sufficiently long relative to the slowest response time
$\tau_r\sim \Delta^{-1}$, which is set by the smallest energy gap
$\Delta$.  Since, as the QCP is approached, the latter vanishes as
$\Delta \sim |\lambda(t)-\lambda_c|^{\nu z}$, adiabaticity is broken
throughout an ``impulse region'' $[t_c -\hat{t}, t_c + \hat{t}]$
symmetrically located around the QCP, where the ``freeze-out'' time
$\hat{t}$ is determined by the condition $\tau(\hat{t})=\tau_r
(\hat{t})$.  In the simplest case of a linear sweep across the QCP
(Fig.  1, top), $\lambda(t)-\lambda_c \equiv t /\tau$ for a fixed rate
$\tau^{-1} >0$, this yields $\hat{t} \sim \tau^{\nu z/(\nu z+1)}$ and
a typical gap $\hat{\Delta} \sim \hat{t}^{-1}$.  Correspondingly, the
typical correlation length $\hat{\xi}\sim \xi (\hat{t}) \sim
\hat{\Delta}^{-z}$ also scales with the quench time $\tau$.  Since
$\hat{\xi}$ is the universal length scale near criticality, it
determines the scaling of the final ($t=t_f$) density of defects and,
more generally, the {\em total} density of excitations,
$n_{\text{ex}}(t_f)$, created in the system.  If $d$ denotes the
spatial dimension, the KZS result then follows:
\begin{eqnarray}
n_{\text{ex}}(t_f) \sim \tau^{-d \nu/(\nu z+1)}.
\label{KZS}
\end{eqnarray}

The validity as well as the limitations of the above KZS have been
carefully scrutinized in a number of settings.  By now, the original
KZS has been confirmed for a variety of models involving a regular
isolated QCP \cite{Jacek,Anatoli,Victor2,Deng1,Sen2}, and extensions
have been introduced for more general adiabatic dynamics, including
repeated \cite{repeated}, non-linear \cite{Sen1}, and optimal
\cite{optimal} quench processes.  In parallel, departures from the KZS
predictions have emerged for more complex adiabatic scenarios,
involving for instance quenches across either an isolated
multicritical point (MCP) \cite{Sen1,Victor,Deng3,Amit,MPD} or
non-isolated QCPs (that is, critical regions)
\cite{Mondal,Pellegrini,Deng2,chowdhury}, as well as QPTs in
infinitely-coordinated \cite{Caneva}, disordered \cite{random}, and/or
spatially inhomogeneous systems\cite{Inhom,JacekRev}.  A main message
that has emerged from the above studies is that, unlike in the
standard KZS of Eq. (\ref{KZS}) where the non-equilibrium critical
exponent is completely specified in terms of static exponents,
additional details about the time-dependent excitation may play an
essential role in general.  As a result, genuinely non-static, {\em
path-dependent exponents} may be required for dynamical scaling
predictions. This feature is vividly exemplified, for instance, in
multicritical quantum quenches, whereby the asymmetry of the KZ
impulse region relative to the static QCP (Fig. 1, bottom) causes a
path-dependent minimum gap other than the critical gap to be relevant
and an effective dynamical exponent $z_2 \ne z$ to emerge
\cite{Deng3}.

In addition to characterizing the response to an adiabatic probe, the
opposite limit of a {\em sudden} change of the tuning parameter
$\lambda(t)$ near a QCP has also attracted a growing attention
recently, in connection with the study of both dynamical
quantum-critical properties \cite{Das,Li,DeGrandi}, as well as
thermalization dynamics in closed quantum systems and its interplay
with quantum integrability
\cite{Cardy,thermalization,Rossini,Sei,Rossini2,Paolo,Canovi,Rigol}.
In particular, for sudden quenches with a sufficiently {\em small
amplitude}, the existence of universal scaling behavior has been
established for various physical observables and qualitatively related
to the above KZ argument \cite{DeGrandi,Gritsev}: by associating to a
quench with amplitude $\delta\lambda$ the characteristic length scale
$\xi \sim |\delta\lambda|^{-\nu}$, and by interpreting $\xi$ as the
correlation length in the final state, one immediately infers that
$n_{\text{ex}}(t_f) \sim \xi^{-d} \sim |\delta\lambda|^{d\nu}$.

\begin{figure}[t]
\centering
\includegraphics[width=7cm]{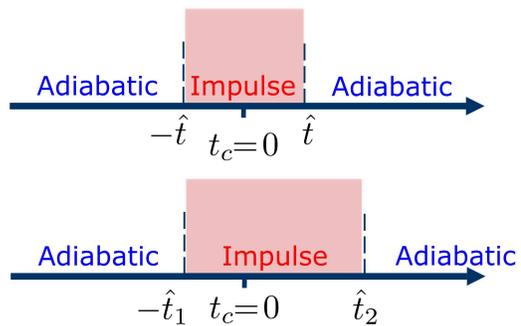}
\vspace{0cm} \caption{(Color online) Qualitative sketch of the
adiabatic-impulse-adiabatic sequence of regimes relevant to dynamical
scaling arguments.  Top: Symmetric impulse region, as assumed by the
standard KZS scenario.  Bottom: Asymmetric impulse region, as
resulting from the existence of quasicritical path-dependent energy
states, see Ref. \onlinecite{Deng3}.}
\label{impulse}
\end{figure}

With a few exceptions where quenches at finite temperature and the
associated thermal corrections have also been examined
\cite{Patane1,DeGrandi,Gritsev,Cardy}, the large majority of the
existing investigations have focused on quench dynamics originating
from the ground state of the initial Hamiltonian $H(t_0)$.  Our goal
in this work is to present a dedicated analysis of {\em finite-energy
quantum quenches}, with a twofold motivation in mind.  Conceptually,
elucidating to what extent and how universal scaling properties may
depend upon the details of the system's initialization is needed to
gain a more complete picture of non-equilibrium quantum-critical
physics.  While one might, for instance, naively expect that a sizable
overlap with the initial ground state would be essential in
determining the applicability of ground-state scaling results, a main
highlight of our analysis is that the support of the initial state on
those excitations relevant to the path-dependent excitation process is
key in a dynamical scenario, in a sense to be made precise later.
Furthermore, from a practical standpoint, perfect initialization of a
many-body Hamiltonian in its exact ground state is both NP-hard in
general \cite{ortiz-somma,poulin} and experimentally unfeasible due to
limited control. In this sense, our investigation both extends
previous studies on finite-temperature signatures of static QPTs
\cite{Sun}, and may be directly relevant to experiments using
ultracold atoms \cite{ExpQPT} as well as nuclear magnetic resonance
(NMR) quantum simulators \cite{negrevergne,Qsim}.

While our analysis focuses on the simplest yet paradigmatic case of an
exactly solvable XY quantum spin chain (Sec. II), we address
non-equilibrium dynamics originating from a large class of (bulk)
initial states for a variety of different quench schemes involving
either a regular QCP or a MCP.  Both pure and mixed initial states
carrying finite excitation energy above the ground state are examined,
under the main assumption that, subsequent to initialization, the
system can be treated as (nearly) isolated, hence evolving under a
time-dependent Hamiltonian.  In particular, dynamical scaling in
adiabatic and sudden quenches starting from an {\em excited energy
eigenstate} are analyzed in Sec. III.A and III.B respectively, with
emphasis on making contact with previously introduced adiabatic
renormalization approaches \cite{Anatoli,Deng2} and on clarifying
formal connections between scaling behavior in sudden and adiabatic
dynamics.  The case of a {\em generic excited pure state} prepared by
a sudden parameter quench is also considered in Sec. III.C, and
criteria are identified for KZS to be obeyed.  Sec. IV is devoted to
quench dynamics resulting from an {\em initial thermal mixture}, with
the main goals of characterizing the robustness of dynamical scaling
behavior in realistic finite-temperature conditions, and of further
exploring the conditions leading to effective thermalization of
certain physical observables following a sudden quench toward
criticality.  In the process, we continue and extend the analysis
undertaken in Deng {\em et al.}\cite{Deng3}, by presenting
finite-temperature generalizations of the scaling predictions obtained
for adiabatic (both linear and non-linear) {\em multicritical quantum
quenches}, as well as evidence of how the peculiar nature of a MCP may
also result in {\em anomalous thermalization behavior}.  Section V
conludes with a discussion of the main findings and their
implications, along with further open problems.

\section{Model Hamiltonian}
\label{General}

\subsection{Energy spectrum and equilibrium phase diagram}

We consider the homogeneous one-dimensional spin-$1/2$ XY model
described by the Hamiltonian:
\begin{eqnarray}
H = - \hspace*{-0.8mm} \sum_{j=1}^N \Big(\frac{1+\gamma}{2}\sigma_x^j
\sigma_x^{j+1} \hspace*{-0.8mm} + \frac{1-\gamma}{2}\sigma_y^j
\sigma_y^{j+1} - \hspace*{-0.7mm} h \sigma_z^j \Big),
\label{Ham}
\end{eqnarray}
where periodic boundary conditions are assumed, that is,
$\sigma_\alpha^j \equiv \sigma_\alpha^{j+N}$, and $N$ is taken to be
even. Here, $\gamma,h \in [-\infty, \infty]$, parameterize the degree
of anisotropy in the XY plane, and the uniform magnetic field
strength, respectively, in suitable units.  The diagonalization of the
Hamiltonian (\ref{Ham}) is
well-known\cite{LSM,Barouch1,Barouch2,Somma}, and we only recall the
basic steps here.  Upon introducing canonical fermionic operators
$\{c_j, c_j^\dagger\}$ via the Jordan-Wigner mapping $c^\dagger_j
\equiv \prod_{\ell=1}^j (-\sigma_z^\ell) \sigma_+^j$, $H$ rewrites as
a quadratic form
\begin{eqnarray}
H &\hspace*{-0.8mm}=\hspace*{-0.8mm}& - \hspace*{-0.8mm}
\sum_{j=1}^{N-1} \,(c_j^\dagger c_{j+1} + \gamma c^\dagger_j
c_{j+1}^\dagger +\text{h.c.}) + 2 h \sum_{j=1}^N c^\dagger_j c_j
\nonumber \\ &\hspace*{-0.8mm}-\hspace*{-0.8mm}& h N + {\cal P}
\,(c^\dagger_N c_1 + \gamma c^\dagger_N c^\dagger_1 +\text{h.c.}),
\label{HamF}
\end{eqnarray}
where the last term originates from the spin periodic boundary
conditions and the parity operator ${\cal P}\equiv \prod_{j=1}^N
(-\sigma_z^j)= e^{i\pi \sum_{j=1}^N c^\dagger_j c_j}=+1 (-1)$
depending on whether the eigenvalue of the total fermionic number
operator is even (odd), respectively. Physically, ${\cal P}$
corresponds to a global ${\mathbb Z}_2$-symmetry which, for finite
$N$, allows the even and odd subspaces to be exactly decoupled,
$H\equiv H^{(+)}+H^{(-)}$, and the diagonalization to be carried out
separately in each sector.

In finite systems, the ground state as well as excited energy
eigenstates with an even number of fermions belong to the ${\cal
P}=+1$ sector.  By using a Fourier transformation to momentum space,
$c_k^\dagger=\frac{1}{ \sqrt{N}}\sum_{j=1}^N e^{-ikj} c_j^\dagger$,
followed by a Bogoliubov rotation to fermionic quasiparticles $\{
\gamma_k, \gamma_k^\dagger\}$, with $\gamma_k = u_k c_k -i v_k
c^\dag_{-k}$, $u_k=u_{-k}$, $v_k=-v_{-k}$, and $u^2_k+v_k^2=1,$ the
Hamiltonian in Eq. (\ref{HamF}) rewrites as a sum of non-interacting
terms:
\begin{eqnarray}
H^{(+)}\equiv \hspace*{-0.8mm}\sum_{k \in K_+} \hspace*{-0.8mm}H_k
=\hspace*{-1mm} \sum_{k \in K_+} \hspace*{-1.5mm} \epsilon_k(h,\gamma)
(\gamma_k^\dag \gamma_{k}+\gamma_{-k}^\dag \gamma_{-k}-1 ).
\label{HamD}
\end{eqnarray}
Here, the set $K\equiv K_++K_-$ of allowed momentum modes is
determined by the anti-periodic boundary conditions on the
fermions in the even sector, $c_{j+N}\equiv -c_j$, which yields
$K_\pm=\Big \{\pm\frac{\pi}{N},\pm\frac{3\pi}{N},\ldots,
\pm\Big(\pi-\frac{\pi}{N}\Big) \Big\}$, and
\begin{eqnarray}
\epsilon_k(h,\gamma)=2\sqrt{(h-\cos{k})^2+\gamma^2\sin{k}^2}
\label{Gap}
\end{eqnarray}
is the quasi-particle energy of mode $k$. For each $k$, let ${\cal
H}_k \equiv \text{span}\{ |0_k\rangle, |1_k\rangle \}$, where
$\{|0_k\rangle, |1_k\rangle = \gamma_k^\dag |0_k\rangle \}$ are
orthonormal states corresponding, respectively, to zero and one
Bogoliubov quasiparticle with momentum $k$, that is, $\langle 0_k
|\gamma_k^\dag \gamma_k |0_k\rangle =0$, $\langle 1_k |\gamma_k^\dag
\gamma_k |1_k\rangle =1$, and similarly for $-k$. Thus, the four
eigenstates of $H_k$ provide a basis for ${\cal H}_k \otimes {\cal
H}_{-k}$,
\begin{eqnarray}
{\cal B}_k & \hspace*{-1mm}= \hspace*{-1mm}& \{ |0_k, 0_{-k}\rangle ,
|1_k,1_{-k}\rangle , |0_k, 1_{-k}\rangle , |1_k, 0_{-k}\rangle \}
\label{basis} \\
& \hspace*{-1mm}\equiv \hspace*{-1mm}&{\cal B}_k^{(+)} \oplus {\cal
B}_k^{(-)} , \nonumber
\end{eqnarray}
where the corresponding eigenenergies are given by $-\epsilon_k$,
$\epsilon_k$, $0$, $0$, and a further separation into even (odd)
sector for each $k$ is possible due to the fact that $[{\cal P}_k
,H_k]=0$, with ${\cal P}_k\equiv e^{i \pi (\gamma_k^\dag \gamma_k
+\gamma^\dag_{-k} \gamma_{-k})} = e^{i \pi (c_k^\dag c_k
+c^\dag_{-k}c_{-k})}$.

The ground state of $H^{(+)}$ corresponds to the BCS state with no
Bogoliubov quasiparticles,
$$ |\Psi_0 ^{(+)}\rangle = \bigotimes_{k\in K_+} |0_k 0_{-k}\rangle =
\bigotimes_{k\in K_+} (u_k +i v_k c^\dag_k c^\dag_{-k})|\mbox{\tt vac}
\rangle, $$
\noindent
where $|\mbox{\tt vac}\rangle$ is the fermionic vacuum.  Many-body
excited states in the even sector can be obtained by applying pairs of
Bogoliubov quasiparticle operators to $|\Psi_0 ^{(+)}\rangle$.  In
particular, excited eigenstates with support only on the even sector
${\cal B}_k^{(+)}$ for each mode are obtained by exciting only pairs
of quasiparticles with opposite momentum and have the form
\begin{eqnarray}
|\Psi^{(+)}_E\rangle=\Big( \bigotimes_{k\in K_+^e} |1_k
1_{-k}\rangle\Big) \Big(\hspace*{-3mm}\bigotimes_{k\in K_+ - K_+^e}
|0_k 0_{-k}\rangle\Big),
\label{evex}
\end{eqnarray}
where $K_+^e$ labels the subset of excited modes.

For finite $N$, the ground state and excited energy eigenstates with
an odd number of fermions belong to the sector ${\cal P}=-1$, which
implies periodic boundary conditions on the fermions, $c_{j+N} \equiv
c_j$, and a different set $\bar{K}$ of allowed momentum modes,
$\bar{K}\equiv \bar{K}_+ + \bar{K}_- + \,\{ 0, -\pi\}$, where
$\bar{K}_\pm = \Big \{ \pm\frac{2\pi}{N},\pm\frac{4\pi}{N},\ldots,
\pm\Big(\pi-\frac{2\pi}{N}\Big) \Big\}$.  Since one may show that
$\epsilon_{k=0}=h-2$ and $\epsilon_{k=-\pi}=h+2$, occupying mode $0$
has always lower energy than occupying mode $-\pi$, thus the ground
state of $H^{(-)}$ is now
$$ |\Psi_0 ^{(-)}\rangle = |1_0 0_{-\pi}\rangle
\bigotimes_{k\in\bar{K}_+} |0_k 0_{-k}\rangle ,$$
\noindent
and excitations may be generated by applying Bogoliubov quasiparticle
operators in such a way that the constraint on the total fermionic
number is obeyed.  Similar to modes in the even sector, $k \in K_+$,
the subspace of each mode $k \in \bar{K}_+$ yields four eigenstates of
$H_k$ and, in principle, a basis formally identical to the one in
Eq. (\ref{basis}) for the odd Hilbert-space sector.  Although for
finite $N$ (thus necessarily in numerical simulations) ${\cal P}$ is
always a good quantum number under dynamics induced by $H$, the error
in the computation of observables arising from identifying the two
sets of modes $K$ and $\bar{K}$ scales like $1/N$.  Thus, for
sufficiently large $N$, a simplified description in terms of a unique
set of momentum modes is possible by using the basis
\begin{equation}
{\cal B}\equiv \bigotimes_{k\in K_+} {\cal B}_k,
\label{basisfull}
\end{equation}
to characterize arbitrary states in the full Hilbert space ${\cal H}=
\bigotimes_{k \in K_+} ({\cal H}_k \otimes {\cal H}_{-k})$.  This
becomes accurate in the thermodynamic limit $N\rightarrow \infty$,
where the many-body ground state becomes twofold degenerate and the
${\mathbb Z}_2$-symmetry spontaneously breaks, causing different
${\cal P}$-sectors to mix.

\begin{figure}[t]
\centering \includegraphics[width=8cm]{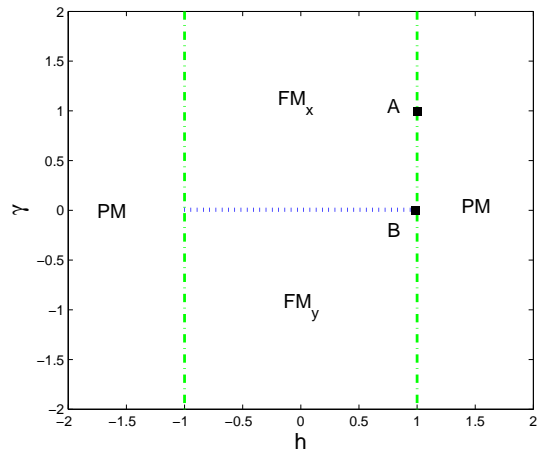}
\caption{(Color online) The phase diagram of the XY Hamiltonian in
Eq.~(\ref{Ham}). The regular QCP {\tt A} ($h_c=1,\gamma_c=1$) and the
MCP {\tt B} ($h_c=1,\gamma_c=0$) are marked. The dashed-dotted (green)
line separates the ferromagnetic (FM) and paramagnetic (PM) phases,
whereas the dotted (blue) line represents the superfluid phase (SF).}
\label{path}
\end{figure}

The equilibrium phase diagram of the model Hamiltonian in
Eq. (\ref{Ham}) is determined by the behavior of the excitation gap of
each mode, $\Delta_k(h,\gamma) \equiv \epsilon_k (h, \gamma)$, with $
\epsilon_k (h, \gamma)$ given in Eq. (\ref{Gap}), and is depicted in
Fig.~\ref{path}.  Throughout this work, we will mainly investigate
scaling behavior in quenches involving either the regular QCP {\tt A}
($h_c=1,\gamma_c=1$), which has equilibrium critical exponents
$\nu=z=1$ and belongs to the $d=2$ Ising universality class, or the
MCP {\tt B} ($h_c=1,\gamma_c=0$), which has $\nu=1/2,z=2$ and belongs
instead to the Lifshitz universality class \cite{Deng3}. In what
follows, we shall refer to the {\em critical mode $k_c$} as the mode
whose gap $\Delta_k(h,\gamma)$ vanishes in the thermodynamic limit.
For both the QCPs {\tt A} and {\tt B} of interest, we thus have
$k_c=0$ in the large-$N$ limit.

\subsection{Dynamical response indicators}

If the system described by Eq. (\ref{Ham}) is driven across a QCP by
making one (or both) of the control parameter(s) $h$, $\gamma$
explicitly time-dependent, excitations are induced as a result of the
non-equilibrium dynamics.  Since the gap vanishes at the QCP in the
thermodynamic limit, this happens no matter how slow the Hamiltonian
changes with time.  In Refs.~\onlinecite{Deng1, Deng2}, the excess
expectation value relative to the instantaneous ground state was shown
to successfully characterize dynamical scaling behavior for a large
class of physical observables in adiabatic quenches originating from
the ground state. That is, for an extensive observable ${\cal O}$, the
following quantity quantifies the underlying adiabaticity loss:
\begin{eqnarray}
\Delta {\cal O}(t) &\equiv & \langle \Psi(t)| \, {\cal O}\,
|\Psi(t)\rangle - \langle \tilde{\Psi}(t) |\, {\cal O}\, |\tilde{\Psi}
(t) \rangle ,
\label{do2}
\end{eqnarray}
where $|\Psi(t)\rangle$ and $|\tilde{\Psi}(t)\rangle$ are the actual
time-evolved state and the adiabatically evolved state resulting from
$|\Psi (t_0) \rangle$, respectively.  For a generic quench process,
where in principle both the time-dependence in $H(t)$ and the initial
state $\rho(t_0)$ can be arbitrary, it is desirable to characterize
the response of the system in such a way that no excitation is
generated by purely adiabatic dynamics \cite{Anatoli} and zero-energy
quenches are included as a special case.  This motivates extending the
definition of Eq. (\ref{do2}) to
\begin{eqnarray}
\Delta {\cal O}(t) &\equiv & \text{Tr}[{\cal O}(t) \rho(t)]-
\text{Tr}[{\cal O}(t) \tilde{\rho}(t)]\,,
\label{do1}
\end{eqnarray}
where now $\rho(t)$ and $\tilde{\rho}(t)$ are the actual time-evolved
density operator and the density operator resulting from adiabatic
evolution of $\rho(t_0)$ respectively, and we also allow, in general,
for the observable ${\cal O}$ to be explicitly time-dependent.  Let
$H(t)|\Psi_i(t)\rangle=E_i(t)|\Psi_i(t)\rangle$ define snapshot
eigenstates and eigenvalues of $H(t)$ along a given control path.
Then the adiabatically steered state
$\tilde{\rho}(t)=\sum_{i,j}\rho_{i,j}(t_0)|
\Psi_i(t)\rangle\langle\Psi_j(t)|$, with $\rho_{i,j}(t_0)$ being the
matrix elements of the initial state $\rho(t_0)$ in the eigenstate
basis $|\Psi_i(t_0)\rangle\langle\Psi_j(t_0)|$ of the initial
Hamiltonian $H(t_0)$.

With respect to the basis given in Eq. (\ref{basisfull}), a generic
uncorrelated state in momentum space may be expressed in the form
$\rho(t)=\bigotimes_{k \in K_+} \rho_k(t)$, where $\rho_k(t)$ is the
four-dimensional density operator for mode $k$.  Relative to a
snapshot eigenbasis
$${\cal B}_k(t)\equiv \{|\psi^{j}_k(t) \rangle \},\quad
j=0,\ldots,3,$$
\noindent
similar to the one given in Eq. (\ref{basis}), but constructed from
time-dependent quasiparticle operators such that $\gamma_k (t)|0_k(t)
\rangle =0$, $\gamma_k^\dag (t)|0_k(t) \rangle =|1_k(t)\rangle$,
$\rho_k(t)$ may be expressed as:
$$\rho_k(t)=\sum_{i,j=0,3}
\rho_{ij,k}(t)|\psi^{i}_k(t)\rangle\langle\psi^j_k(t)|.$$
\noindent
Suppose that the time-evolution operator for mode $k$ is $U_k(t)$,
that is, $\rho_k(t)=U_k(t) \rho_k(t_0) U^\dag_k(t)$.  Direct
calculation shows that $|0_k(t),1_{-k}(t)\rangle=c^\dag_{-k}|{\tt
vac}\rangle$, and $|1_k(t),0_{-k}(t)\rangle=c^\dag_{k}|{\tt
vac}\rangle$ for all $t$, which indicates that the snapshot
eigenstates belonging to the ${\cal P}_k=-1$ eigenvalues are frozen in
time, $|0_k(t),1_{-k}(t)\rangle=|0_k(t_0),1_{-k}(t_0)\rangle, \;\;\;
|1_k(t),0_{-k}(t)\rangle=|1_k(t_0),0_{-k}(t_0)\rangle.$ As long as
${\cal P}_k$ is conserved under $H_k(t)$, the even and odd sectors for
each $k$ are decoupled.  Thus, upon letting
\begin{eqnarray*}
U_k^\dag(t)|1_k(t),1_{-k}(t)\rangle & \equiv & a_{0,k}(t)|0_k
(t_0),0_{-k}(t_0)\rangle \\ &+ &
a_{1,k}(t)|1_k(t_0),1_{-k}(t_0)\rangle,
\end{eqnarray*}
we can evaluate the time-dependent excitation probability for mode $k$
as follows:
\begin{eqnarray}
\hspace{-10mm}P_k(t)\hspace{-1mm}& \equiv & \hspace{-1mm}
\text{Tr}[\rho_k(t) \gamma_k^\dag(t) \gamma_k(t)]\nonumber \\
\hspace{-1mm}&=& \hspace{-1mm}
\text{Tr}[\rho_k(t)(|1_k(t),1_{-k}(t) \rangle\langle 1_k(t),1_{-k}(t)|]
\nonumber \\
\hspace{-1mm}&+& \hspace{-1mm}\text{Tr}[\rho_k(t)|1_k(t),0_{-k}(t)
\rangle\langle 1_k(t),0_{-k}(t)|]\nonumber \\
\hspace{-1mm}&=&\hspace{-1mm}
(\rho_{00,k}(t_0)-\rho_{11,k}(t_0))|a_{0,k}(t)|^2
+\rho_{11,k}(t_0) \nonumber
\\ \hspace{-1mm}&+&\hspace{-1mm} 2 \text{Re}[\rho_{01,k}(t_0)
a_{0,k}^\ast(t)a_{1,k}(t)]+\rho_{33,k}(t_0),
\label{rhot}
\end{eqnarray}
where the relationships $|a_{0,k}(t)|^2+|a_{1,k}(t)|^2=1$ and
$\rho_{10,k}=\rho_{01,k}^\ast$ have been exploited. Notice that from
the above definition of $a_{0,k}(t)$, it follows that $|a_{0,k}(t)|^2$
is the time-dependent probability that mode $k$ is excited when it is
in its ground state at $t=t_0$. Similarly, we may express the
adiabatically evolved density operator
$\tilde{\rho}(t)=\bigotimes_{k\in K_+} \tilde{\rho}_k(t)$, with
$\tilde\rho_k(t)=\sum_{i,j=0,3}
\rho_{ij,k}(t_0)|\psi^{i}_k(t)\rangle\langle\psi^j_k(t)|$.  Thus, the
time-dependent excitation probability of mode $k$ relative to the
adiabatic path is simply
\begin{eqnarray}
\tilde{P}_k(t) & \hspace{-1mm}=\hspace{-1mm}&
\text{Tr}[\tilde{\rho}_k(t) \gamma_k^\dag(t) \gamma_k(t)] \nonumber \\
&\hspace{-1mm} = \hspace{-1mm}&
\rho_{11,k}(t_0)+\rho_{33,k}(t_0)\equiv P_k(t_0).
\label{rhotad}
\end{eqnarray}
Upon combining Eqs. (\ref{rhot})-(\ref{rhotad}), the {\em relative
excitation probability} of mode $k$ is given by
\begin{eqnarray}
\Delta P_k(t) &\hspace{-1mm}\equiv & \hspace{-1mm} P_k(t)-P_k(t_0)
\nonumber \\ &\hspace{-1mm} =\hspace{-1mm} &
(\rho_{00,k}(t_0)-\rho_{11,k}(t_0)) |a_{0,k}(t)|^2 \nonumber \\
&\hspace{-1mm}+ \hspace{-1mm} & 2 \text{Re}[\rho_{01,k}(t_0)
a_{0,k}^\ast(t)a_{1,k}(t)].
\label{relativex}
\end{eqnarray}

Physically, a non-zero contribution $P_k(t_0)$ may account for initial
excitations due to either a coherent preparation into an excited state
or to a finite temperature $T$.  Clearly, if mode $k$ is initially in
its ground state, $P_k(t_0)=0$, we consistently recover the
definitions in Ref.~\onlinecite{Deng2} for zero-energy quenches. Two
relevant limiting cases of Eq. (\ref{relativex}) will play a special
role in what follows.  First, if mode $k$ is initially in a generic
pure state of the form $$|\psi_{k}(t_0)\rangle \equiv
\sum_{j=0,3}c_{j,k}|\psi^j_k(t_0)\rangle,$$ then $\rho_{00,k}(t_0)=
|c_{0,k}|^2$, $\rho_{01,k}(t_0)= c_{0,k}c_{1,k}^\ast$,
$\rho_{11,k}(t_0)=|c_{1,k}|^2$, hence
\begin{eqnarray}
\Delta P_k(t)&=&(|c_{0,k}|^2-|c_{1,k}|^2)|a_{0,k}(t)|^2 \nonumber \\
&+& 2\text{Re}[c_{0,k}c_{1,k}^\ast a_{0,k}^\ast(t)a_{1,k}(t)].
\label{Pexpure}
\end{eqnarray}
Second, if the initial state $\rho(t_0)$ is a statistical mixture,
then $\rho_{10,k}(t_0)=\rho_{01,k}(t_0)=0$, and we have
\begin{eqnarray}
\Delta P_k(t)=(\rho_{00,k}(t_0)-\rho_{11,k}(t_0))|a_{0,k}(t)|^2.
\label{Pexmixed}
\end{eqnarray}

The time-dependent excess expectation value $\Delta {\cal O}(t)$ in
Eq. (\ref{do1}) may be expressed directly in terms of the relative
excitation probability for observables that obey $[{\cal O}(t),
H(t)]=0$ at all times. In this work, we shall primarily focus on the 
following choices:

$\bullet$ ${\cal O}(t)= \frac{1}{N}\sum_{k\in {K^+}} [\gamma_k^\dag
(t) \gamma_k (t) + \gamma_{-k}^\dag (t)\gamma_{-k}(t)]$, leading to
the relative total density of excitations: density:
\begin{eqnarray}
\Delta n_{\text{ex}}(t) &=&\frac{2}{N}\sum_{k \in
K_+}\text{Tr}[(\rho_k(t)-\tilde{\rho}_k(t)) \gamma_k^\dag(t)
\gamma_k(t)]\nonumber \\ &=&\frac{2}{N} \sum_{k \in K_+} \Delta
P_{k}(t),
\label{dnex}
\end{eqnarray}
which coincides with $n_{\text{ex}}(t)$ when the initial state is the
ground state.

$\bullet$ ${\cal O}(t) = H(t)$, leading to the relative excitation
energy density:
\begin{eqnarray}
\Delta H(t) &=&\frac{2}{N}\sum_{k \in K_+}
\text{Tr}[(\rho_k(t)-\tilde{\rho}_k(t))H_k(t)]\nonumber \\
&=&\frac{2}{N} \sum_{k \in K_+} \epsilon_k(h(t),\gamma(t))\Delta
P_{k}(t).
\label{dH}
\end{eqnarray}

\noindent
While $\Delta n_{\text{ex}}(t)$ is especially attractive from a theory
standpoint in view of its simplicity (possibly enabling analytical
calculations), a potential advantage of $\Delta H(t)$ is that it may
be more directly accessible in experiments.

As a representative example of an observable {\em not} commuting with
the system's Hamiltonian, we shall additionally include results on the
scaling behavior of:

$\bullet$ ${\cal O}\equiv XX= \frac{1}{N} \sum_{i=1}^N \sigma_x^i
\sigma_x^{i+1}$, corresponding to the nearest-neighbor spin correlator
per site along the $x$-direction \cite{Deng1}.  In the Ising limit
($\gamma=1$), the operator ${\cal N}\equiv (1-XX)/2$ is a natural
measure for the ``density of kinks'' created by the quench, which
directly relates to the number of quasi-particles excited at $h=0$
\cite{Jacek,JacekRev,Rossini2}.  We have:
\begin{eqnarray}
\Delta XX(t) &\hspace*{-1mm}=\hspace*{-1mm}&
\frac{2}{N}\hspace{-1mm}\sum_{k \in
K_+}\Delta\big{(}\hspace{-1mm}-\hspace{-1mm}2\cos{k}\, c_k^\dag
c_k \big) \nonumber \\ &\hspace*{-1mm}+\hspace*{-1mm}& \Delta\big(
i \gamma(t) \sin{k}\,(c^\dag_{k}c^\dag_{-k}- \text{h.c.})\big{)}
\label{dxx}
\end{eqnarray}

In principle, the sums in Eqs.~(\ref{dnex})--(\ref{dxx}) should
include all the modes in $K_+$, as indicated. However, to the purpose
of analytically investigating dynamical scaling behavior, it is useful
to note that not all the allowed modes will necessarily change their
state along the adiabatic quench path, effectively making no
contribution to the relative expectation $\Delta {\cal O}(t)$.  In
what follows, we shall refer to the subset of modes $K_R \subseteq
K_+$ whose state changes in an adiabatic quench as the {\em relevant
modes}.  Let a power-law adiabatic quench process be parametrized as
$\lambda(t)- \lambda_c = |t/\tau|^\alpha \text{sign}(t)$, $\alpha =1$
corresponding to the standard linear case also discussed in the
Introduction. We may relate the number of relevant modes, $N_R\equiv
|K_R|$, to the system size and the quench rate via $N_R(N,\tau)\propto
N |k_{\text{max}}(\tau)-k_c|$, where $k_{\text{max}}$ is the largest
momentum in the relevant mode set.  Since adiabaticity breaks at a
time scale $\hat{t}\sim \tau^{\alpha \nu z/(\alpha \nu z+1)}$, and the
typical gap, $\hat{\Delta} \sim \hat{t}^{-1}$, an accessible excited
state contributes to the excitation if and only if its minimum gap
along the path, $\tilde{\Delta}_k$, matches with this typical gap,
$\tilde{\Delta}_k \sim \hat{\Delta}$.  In general \cite{Deng3},
$\tilde{\Delta}_k$ scales as $\tilde{\Delta}_k \sim (k-k_c)^{z_2}$,
where $z_2$ is a genuinely non-static exponent.  Accordingly, the
scaling of $k_{\text{max}}$ can be determined by $\hat{\Delta} \sim
(k_{\text{max}}-k_c)^{z_2}$, leading to
\begin{equation}
k_{\text{max}}-k_c \sim \tau^{-{\alpha\nu z}/{[z_2(\alpha\nu
z+1)]}}.
\label{kmax}
\end{equation}

\section{Quenches from a pure excited state}
\label{ExcEig}

\subsection{Adiabatic quench dynamics from an excited energy eigenstate}

Adiabatic quenches from the ground state of the initial Hamiltonian
$H(t_0)$ have been extensively studied and are well understood in this
model \cite{Jacek,Anatoli,Victor2,Deng1,Deng3}.  In order to explore
the role of initialization, a first natural step is to investigate
dynamical scaling behavior when the initial state is an excited
eigenstate of $H(t_0)$.  Since, as remarked, the time-evolution of
excited components along $|0_k,1_{-k}\rangle$ and $|1_k,0_{-k}\rangle$
is trivial, we shall focus on excited energy eigenstates with support
only on the even sector of each mode $k$, that is, on states of the
form given in Eq. (\ref{evex}).  Noting that there are only two
possibilities for each mode, either $c_{0,k}=1$ or $c_{0,k}=0$, Eq.
(\ref{Pexpure}) yields
\begin{eqnarray}
\Delta P_k(t)=(|c_{0,k}|^2-|c_{1,k}|^2)|a_{0,k}(t)|^2=\pm
|a_{0,k}(t)|^2,\;\;
\label{eiex}
\end{eqnarray}
and, correspondingly,
\begin{eqnarray}
\Delta n_{\text{ex}}(t_f)= \frac{2}{N} \sum_{k \in K_R} \Big[\pm
  |a_{0,k}(t_f)|^2 \Big].
\label{eiexn}
\end{eqnarray}
Thus, the relative excitation density is the same, up to a sign, in
two limiting cases: (i) the many-body ground state, corresponding to
$c_{0,k}=1$ for all $k$ and to an overall positive sign in
Eq.~(\ref{eiexn}); and (ii) the state where all allowed pairs of
quasiparticles are excited, corresponding to $c_{0,k}=0$ for all $k$
and to an overall negative sign in Eq.~(\ref{eiexn}).  Since KZS is
known to hold for a linear quench process with initial condition (i),
and a global sign difference would not change the scaling behavior,
KZS is expected to persist for the maximally excited initial
eigenstate (ii) as well.  This is to some extent surprising both in
view of the fact that such an initial state has zero overlap with the
BCS state $|\Psi_0^{(+)}\rangle$, and because one would not {\em a
priori} expect highly energetic eigenstates to be sensitive to the
ground-state QPT.

Interestingly, critical properties of excited eigenstates in the XY
chain have recently attracted attention in the context of {\em static}
QPTs~\cite{Alba}.  Suppose that each excited eigenstate is associated
with an ordered binary strings of length $|K|= N$, where $0$ ($1$)
represents a mode in its ground (excited) state, respectively.  Then a
compact description of the eigenstate may be given in terms of the
``discontinuities'' of a (suitably regularized as $N\rightarrow
\infty$) characteristic function of the corresponding occupied mode
set, where no discontinuity is present when all modes are $0$ or $1$,
and a discontinuity is counted every time the occupation of a given
mode changes along the string.  Alba {\em et al.}~\cite{Alba} have
analytically proved, in particular, that the block entanglement
entropy, $S_L$, of an excited eigenstate of the critical XY chain may
still obey conformal scaling as in the ground state provided that the
number of discontinuities remains {\em finite} in the thermodynamic
limit, that is, $S_L \sim \log L$, where $L \gg 1$ is the block size.
Conversely, $S_L$ exhibits non-critical scaling, $S_L \sim L$, when
the number of discontinuities becomes itself an extensive quantity as
$N\rightarrow \infty$. Thus, certain highly excited states (including
the fully excited state considered above) can still display critical
behavior, the number of discontinuities in the {\em full} set of
momentum modes being the essential factor in determining the {\em
static} scaling behavior.  While, intuitively, non-analyticities in
the characteristic mode occupation function need not play a direct
role for simple observables such as the excitation density, these
results still prompt the following question: to what extent does a
distinction between ``critical'' (leading to KZS) and ``non-critical''
excited eigenstates exist for {\em dynamical} QPTs?

\begin{figure}[t]
\centering
\includegraphics[width=9cm]{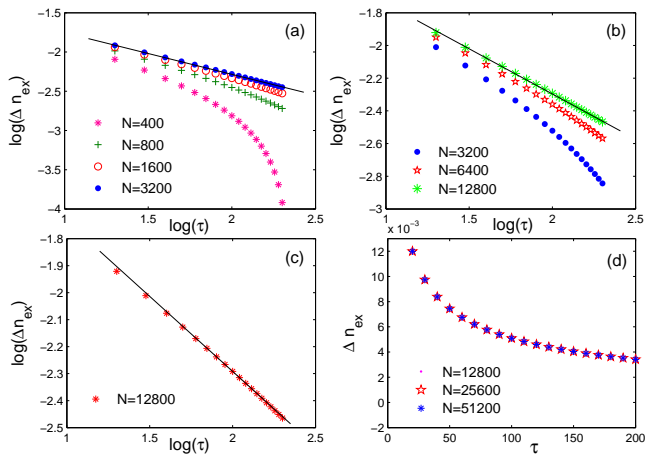}
\caption{(Color online) Scaling behavior of the final relative
excitation density in a linear quench of the magnetic field $h$ around
the QCP {\tt A} in the Ising chain, starting with different excited
eigenstates of $H(h_c)$. Panel (a): only $k_c=\pi/N$ is excited
initially.  The linear fit for $N=3200$ yields $-0.535 \pm
0.002$. Panel (b): the five lowest-energy modes are initially excited.
A linear fitting slope of $-0.549 \pm 0.003$ is now reached at
$N=12800$.  Panel (c): five modes ($k=k_c$, $5\pi/N$, $9\pi/N$,
$13\pi/N$, and $17\pi/N$) are initially excited. The linear fit for
$N=12800$ yields $-0.546 \pm 0.002$. Panel (d): the five lowest-energy
modes are initially excited for $N=12800$ as in (b), but as the system
size is increased linearly, the number of excited modes is increased
accordingly.  In all cases, the relevant $\tau$-range
$\tau_{\text{min}}< 20 \leq \tau \leq 250 < \tau_{\text{max}}$ (see
text and Fig. \ref{extra}). }
\label{fig3}
\end{figure}

\begin{figure}[t]
\centering
\includegraphics[width=8cm]{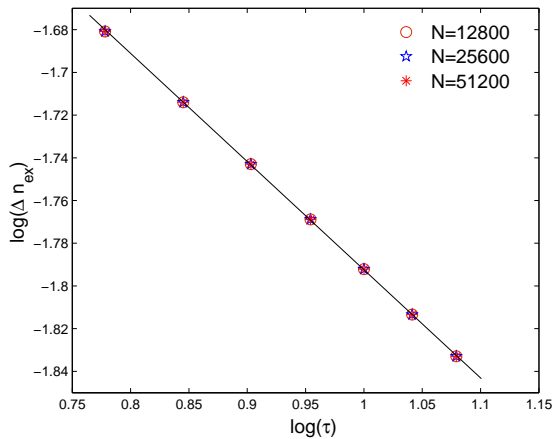}
\caption{(Color online) Scaling behavior of the final relative
excitation density in a linear magnetic-field quench around the QCP
{\tt A} in the Ising chain, starting with an eigenstate of $H(t_c)$
where the five, ten, and twenty lowest-energy modes are initially
excited for $N=12800, N=25600, N=51200$, respectively [same as in
Fig. \ref{fig3}(d)].  The relevant $\tau$-range is now
$\tilde{\tau}_{\text{min}} \sim \tau_{\text{min}} =5 \leq \tau \leq
\tilde{\tau}_{\text{max}}=20 \ll \tau_\text{max}$.  A linear fitting
slope of $-0.5019 \pm 0.002$ is now reached for all these cases, in
agreement with the KZS prediction.} 
\label{extra}
\end{figure}

A key difference with respect to the static situation is that only the
{\em relevant} modes matter in a dynamical QPT, $k \in K_R \equiv
[k_c,k_{\text{max}}]$, with $k_{\text{max}}$ given in
Eq. (\ref{kmax}).  In Fig.~\ref{fig3}, we present exact numerical
results, obtained by direct numerical integration of the
time-dependent Schr\"odinger equation, for the relative excitation
density in a linear adiabatic quench of the magnetic field $h$ around
the QCP {\tt A} of the critical Ising chain ($\gamma=1$). Different
initial eigenstates are compared over a common range of $\tau$, which
is chosen to be well within the appropriate $\tau$-range \cite{Deng2}
for ground-state quenches (see next paragraph and Fig. \ref{extra} for
further discussion of this point). In panel (a), the system is
initialized in the first excited state, where only the critical mode
is initially excited (thus only one discontinuity is present), whereas
in panel (b), the five lowest-energy modes are initially excited
(leading to one discontinuity as well).  In case (a), while no scaling
is visible for a system with size $N=400$, progressively better
scaling behavior emerges as $N$ is increased, with the value at
$N=3200$ approaching the asymptotic KZS value (and better agreement
being achievable by optimizing the $\tau$-range, see below).  In
constrast, for the data in panel (b), a system size as large as
$N=12800$ is required for a scaling of comparable quality to be
established.  Since the only difference between cases (a) and (b) is a
different (fixed) number of initially excited modes in $N_R$, the fact
that upon increasing $N$ (thereby increasing $N_R$ accordingly) a
better KZS is comparatively obtained in (a) suggests that the {\em
ratio between the number of initially excited (or non-excited) modes
and $N_R$} is crucial for dynamical scaling behavior -- not (as
intuitively expected) the discontinuity properties which characterize
the initial mode occupation {\em per se}. More explicitly, let $N_E$
denote the number of modes in $K_R$ that are excited at time $t_0$,
with $N_R-N_E$ correspondingly denoting the number of non-excited
modes in $K_R$, and let
$$ M_R \equiv \min \{N_E, N_R-N_E \}.$$
\noindent
Motivated by the above observations and also recalling the symmetric
role played by initially non-excited vs. excited modes in determining
the time-dependent relative probability of excitation
[Eq.~(\ref{eiex})], we conjecture that KZS emerges in the
thermodynamic limit provided that the initial excited eigenstate
satisfies
\begin{eqnarray}
\frac{M_R}{N_R} =\varepsilon \ll 1.
\label{condition}
\end{eqnarray}
\noindent
Clearly, the case of ground-state initialization corresponds to
$N_E=M_R=0$, and the fully excited state coincides with $N_E=N_R,
M_R=0$.  For a generic initial excited eigenstate,
Eq. (\ref{condition}) allows in principle $M_R$ to be an extensive
quantity in the thermodynamic limit.  Two additional results are
included in Fig.  \ref{fig3} to illustrate the above possibility.  In
panel (c), we still have five excited modes in $N_R$ as in (b), but
five discontinuities as opposed to just one. For the same system size
(thus also the same $\varepsilon$), the scaling is not worse than in
panel (b), further supporting the conclusion that the number of
discontinuities does not play a role towards the emergence of
dynamical scaling.  In panel (d), a fixed value of $\varepsilon$,
equal to the one in (b) at $N=12800$, is explored for different values
of $N$, by also proportionally increasing $N_E$. As the data show, the
resulting $\Delta n_{\text{ex}}$ is the same, indicating that $M_R$
may indeed be allowed to be an extensive quantity as long as Eq.
(\ref{condition}) is obeyed.

It is important to address how the choice of a range of $\tau$-values
affects the above scaling conclusions.  Let $\tau_{\text{min}} \leq
\tau \leq \tau_{\text{max}}$ and $\tilde{\tau}_{\text{min}} \leq \tau
\leq \tilde{\tau}_{\text{max}}$ denote the valid range for
ground-state initialization \cite{Deng2}, and for excited-state
initialization, respectively.  Since $\tau_{\text{min}}$ is determined
from the requirement that an adiabatic regime exists away from
criticality, whereas $\tau_{\text{max}}$ follows from ensuring that
adiabaticity can be broken in a finite-size system, both
$\tau_{\text{min}}$ and $\tau_{\text{max}}$ are related to the scaling
of the {\em many-body} gap between the ground state and first
(available) excited state.  Thus, $\tilde{\tau}_{\text{min}}$
($\tilde{\tau}_{\text{max}}$) could in general be substantially
different from $\tau_{\text{min}}$ ($\tau_{\text{max}}$),
respectively.  In our case, however, the Hamiltonian in
Eq.~(\ref{Ham}) can be exactly decoupled into two-level systems for
each mode $k$. Therefore, the relevant gap is always $\Delta_
{k_c}\equiv \varepsilon_{k_c}(h,\gamma)$, {\em irrespective of the
initial condition}. For this reason, the relation $\tau_{\text{min}}
\leq \tilde{\tau}_{\text{min}}$ and $\tilde{\tau}_{\text{max}} \leq
\tau_{\text{max}}$ must hold, as any finite-energy initial state might
imply more restrictive constraints as compared to the zero-energy
case. In particular, according to Eq. (\ref{condition}), not all the
excited eigenstates can lead to KZS, and the better
Eq.~(\ref{condition}) is satisfied, the closer KZS will be
approached. This explains why, for instance, the fitting slope
$-0.549$ from panel (b) of Fig.~\ref{fig3} is not as close to the KZ
value as the one obtained for a ground-state quench with the same
range of $\tau$.  In the setting of Fig.~\ref{fig3}(b), the majority
of the relevant modes stay in their ground state. In order to reduce
the contribution to $\Delta n_{\text{ex}}$ from the five lowest-energy
modes, we can decrease $\tau$ such that $N_R$ will be increased in
Eq. (\ref{condition}). Numerical support for this strategy is shown in
Fig.~\ref{extra}, where an optimal range of $\tau$ is identified for
the same initial states as in Fig.~\ref{fig3}(d), and very good
agreement with KZS is recovered.  Thus, we can conjecture that if the
majority of modes that enter $M_R$ are low-energy modes, we can reduce
their contribution to $\Delta n_{\text{ex}}$ by decreasing the upper
bound to $\tau$. That is, we choose $\tau_{\text{min}} \leq \tau \leq
\tilde{\tau}_{\text{max}}$, where $\tilde{\tau}_{\text{max}} <
\tau_{\text{max}}$.  Conversely, if the majority of modes that are
enter $M_R$ are from high-energy modes, then we reduce the
contribution from these modes by increasing the lower bound to
$\tau$. That is, we let $\tilde{\tau}_{\text{min}} \leq \tau \leq
\tau_{\text{max}}$, with $\tilde{\tau}_{\text{min}} >
\tau_{\text{min}}$.

Additional theoretical understanding of the criterion given in
Eq. (\ref{condition}) may be sought by invoking the perturbative
Adiabatic Renormalization (AR) approach \cite{Messiah}, which was
successfully applied to explain the scaling results for adiabatic
quenches starting from the ground state \cite{Anatoli,Deng2}.  Can
first-order AR still capture dynamical scaling for initial excited
eigenstates?  Let us focus on linear quenches ($\alpha=1$), and let
the time-dependent Hamiltonian be parametrized as
$H(t)=H_c+[\lambda(t)-\lambda_c] H_1= H_c + (t/\tau) H_1$, with $H_c$
quantum-critical in the thermodynamic limit, so that the relevant QCP
is crossed at $t_c\equiv 0$. If the system is prepared in the
$\ell$-th eigestate of $H(t_0)$, with $t_0 \rightarrow t_c$ as in the
examples previously considered, the time-evolved state from
first-order AR may be expressed in the form
$$|\Psi^{(1)}(t)\rangle=e^{-i\Gamma_\ell(t)}|\Psi_\ell(t)\rangle
-\sum_{m\neq \ell} c_m^{(1)}(t)|\Psi_m(t)\rangle, $$
\noindent
where $\Gamma_\ell(t)$ includes in general both the Berry phase and
the dynamical phase, and $c_m^{(1)}(t)$ gives the time-dependent
amplitude along the $m$-th snapshot eigenstate.  Following a
derivation similar to the one given in Refs.
\onlinecite{Messiah,Deng2}, and letting $\Delta_m(t) \equiv
E_m(t)-E_\ell(t)$, we find:
\begin{eqnarray*}
\hspace*{-3mm}c_m^{(1)} (t)&\hspace*{-1.0mm}=\hspace*{-1.0mm}&
\frac{e^{-i\Gamma\hspace*{-0.5mm}_m\hspace*{-0.5mm}(t)} }{\tau}
\hspace*{-1.0mm}\int_{t_{0}}^t \hspace*{-1.0mm} dt'
\frac{\langle
\Psi_m(t')|H_1|\Psi_\ell(t')\rangle}{E_m(t')-E_\ell(t')}\hspace*{0.5mm}
e^{i\int_{t_{0}}^{t'} \hspace*{-1.0mm} ds \Delta_m(s) }.
\end{eqnarray*}
Thus, to first order in the quench rate $1/\tau$ the adiabaticity loss
can be quantified by $\Delta {\cal O}(t)=\langle\Psi^{(1)}(t)|{\cal
O}(t)|\Psi^{(1)}(t)\rangle-\langle\Psi_\ell(t)|{\cal
O}(t)|\Psi_\ell(t)\rangle$. In particular, this yields
\begin{eqnarray*}
\Delta n_{\text{ex}}(t)
&\hspace*{-1.5mm}=\hspace*{-1.5mm}&\frac{2}{N}\hspace*{-1.0mm}\sum_{m
\neq \ell} |c_m^{(1)}(t)|^2
\Big(\langle\Psi_m(t)|\hspace*{-1.0mm}\sum_{k\in K_+} \gamma^\dag_k
(t) \gamma_k (t)|\Psi_m(t)\rangle \nonumber \\
&\hspace*{-1.5mm}-\hspace*{-1.5mm}&\langle\Psi_\ell(t)|\hspace*{-1.0mm}
\sum_{k\in K_+}\gamma^\dag_k (t) \gamma_k
(t)|\Psi_\ell(t)\rangle\Big).
\end{eqnarray*}
\noindent
Since $H_1$ is a one-body operator in our case, the only non-zero
matrix elements $\langle\Psi_m(t)|H_1|\Psi_\ell(t)\rangle$ include
many-body eigenstates $|\Psi_m(t)\rangle$ which differ from
$|\Psi_\ell(t)\rangle$ in the {\em occupation of precisely one mode}.
Thus, $\langle\Psi_m(t)|\sum_k\gamma^\dag_k (t) \gamma_k (t)
|\Psi_m(t)\rangle-\langle\Psi_\ell(t)|\sum_k\gamma^\dag_k (t) \gamma_k
(t) |\Psi_\ell(t)\rangle=\pm 1$, which implies
\begin{equation}
\Delta n_{\text{ex}}(t_f)=\frac{1}{N} \sum_{m \neq \ell} \Big [\pm
  |c_m^{(1)}(t_f)|^2 \Big] .
\label{ar}
\end{equation}
Except for a possible sign difference for each $m$, the above
expression is formally identical to the one holding for ground-state
initialization ($\ell=0$), in analogy with the exact Eq.
(\ref{eiexn}). Numerical calculations of the relative excitation
density according to Eq. (\ref{ar}) [for instance with the same
initial condition as in Fig.~\ref{fig3}(a), data not shown] confirm
that the condition for initial excited eigenstates to support KZS is
the same in first-order AR as the one conjectured based on exact
numerical results.

\subsection{Sudden quench dynamics from an excited energy eigenstate}
\label{Suddenex}

As mentioned in the Introduction, scaling results for sudden
quenches of the control parameter $\lambda$ around its critical
value $\lambda_c$ have been recently obtained by De Grandi {\em et
al.}~\cite{DeGrandi} under the assumptions that the system is in
the ground state of the initial Hamiltonian and the quench has a
{\em small amplitude}, leading to a final excitation density
\begin{equation}
n_{\text{ex}} (t_f) \sim |\lambda -\lambda_c|^{d \nu} \equiv
\delta\lambda^{d \nu},
\label{sudden}
\end{equation}
with $\delta\lambda \ll 1$ in suitable units. Before addressing, in
analogy to the case of adiabatic dynamics, the extent to which the
expected scaling behavior may be robust against initialization in a
finite-energy eigenstate, it is useful to explore more quantitatively
the connection between ground-state adiabatic vs.  sudden quenches
implied by Eq. (\ref{sudden}).

\begin{figure}[tb]
\includegraphics[width=8cm]{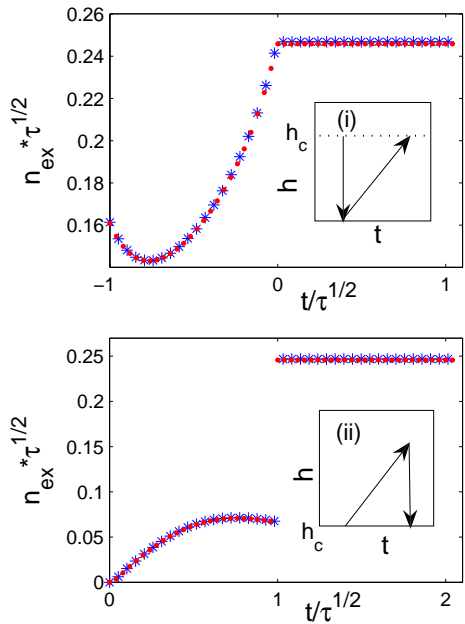}
\caption{(Color online): Scaling behavior of the final excitation
density in combined magnetic-field ground-state quenches across QCP
{\tt A} in the Ising chain.  Top: Sudden quench $h_c \mapsto h_f$ (see
text) followed by a linear quench back to $h_c$, with the system
finally kept at $h_c$.  Bottom: Linear quench from $h_c$ followed by a
sudden quench $h_f \mapsto h_c$, with the system finally kept at
$h_c$. In both cases, $N=400$.  }
\label{fig4}
\end{figure}

\begin{figure}[th]
\centering
\includegraphics[width=8cm]{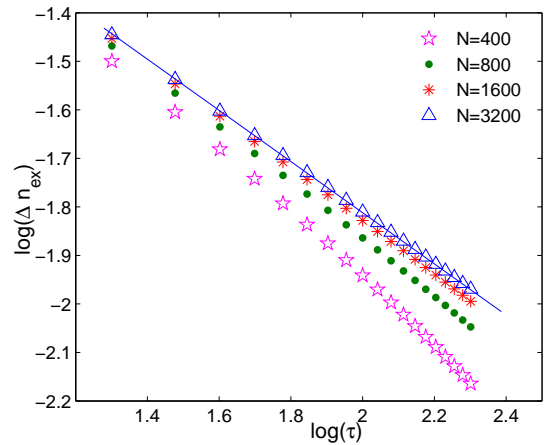}
\caption{(Color online) Scaling behavior of the final relative
excitation density in a sudden magnetic-field quench across QCP {\tt
A} in the Ising chain, starting with the first excited state of
$H(h_c)$. The linear fitting slope for $N=3200$ is $-0.5244 \pm
0.0004$ for $20 \leq \tau\leq 250$.  Closer agreement with the KZS may
be reached by optimizing over $\tau$ as in Fig. \ref{extra}.}
\label{fig5}
\end{figure}

Suppose, specifically, that the amplitude of a sudden magnetic-field
quench near the QCP {\tt A} of the Ising chain is directly related to
the rate $\tau$ of a corresponding linear adiabatic sweep across the
same QCP via $h_f=h_c\pm\hat{h}$, where $\hat{h} \propto \hat{t}/\tau$
and $\hat{t}$ is the KZ freeze-out time scale, {\em i.e.}, $\hat{t}
\sim \tau^{\nu z/(\nu z+1)}$. Equation (\ref{sudden}) then yields
\begin{equation}
n_{\text{ex}} (t_f) \sim |h_f-h_c|^{d \nu} \sim \tau^{-d \nu/(\nu
z+1)}.
\label{suddentau}
\end{equation}
In other words, the scaling behavior resulting from Eq.
(\ref{sudden}) is essentially equivalent to KZS.  While this could be
quantitatively demonstrated by direct calculation of $n_{\text{ex}}
(t)$ in a sudden quench, it can also be nicely illustrated by
examining {\em combined} sudden-adiabatic quenches, which have not
been explicitly addressed to our knowledge \cite{NoteDecoh}, and will
also be relevant in Sec. \ref{ExcSt}.  Two possible ``control loops''
starting from $h(t_0)=h_c$ are depicted in Fig.~\ref{fig4}: we can
either (i) suddenly change the magnetic field amplitude $h_c \mapsto
h_f$, and then adiabatically change it back to $h_c$ (top panel); or
we can (ii) slowly ramp up $h_c$ to $h_f$, and then suddenly quench
$h_f \mapsto h_c$ (bottom panel). As it is clear from the numerical
data, the total excitation density created from the combined
sudden-adiabatic quench shows KZS throughout the entire process in
both cases, provided that $\tau$ is within the appropriate scaling
range ${\tau}_{\text{min}} \leq \tau \leq \tau_{\text{max}}$.  Notice
that the quench process depicted in Fig.~\ref{fig4} is similar to the
repeated linear quench across QCP {\tt A} studied in Ref.
\onlinecite{repeated}, in the sense that the initial and final value
of the control parameter coincide. While KZS was found to hold in such
a repeated linear quench, the difference is now that half of the
linear adiabatic sweep is replaced by a sudden quench.  Since,
however, the interval $[h_c-\hat{h},h_c+\hat{h}]$ corresponds to the
impulse region in the KZ scenario for a pure linear quench, the
scaling results of the combined quenches under consideration may be
understood as a consequence of the fact that the sudden quench
component can only further {\em enforce} the impulse mechanism by
which excitation is generated in the KZS argument.  Interestingly, as
long as the scaling exists, we can also observe that (i) and (ii) lead
to almost the same final excitation density, even if the intermediate
values of the excitation density after the sole sudden [in (i)] or
linear [in (ii)] quench are different.  In summary, the existence of
KZS in ground-state sudden and combined sudden-adiabatic quenches with
{\em small amplitude} is essentially a reflection of the fact that the
system goes through an impulse region around the QCP no matter how
slow or fast the quench is effected.

While sudden quenches of {\em arbitrary amplitude} will be further
considered in the next Section, we now return to the question of
whether dynamical scaling also holds in small-amplitude sudden
quenches when the system is initially prepared in an excited
eigenstate of $H(t_0)=H_c$.  Exact numerical results are presented in
Fig.~\ref{fig5}, where in order to ease the comparison with a linear
quench, we have again explicitly related the sudden-quench amplitude
to $\tau$ as $h_f-h_c \propto \tau^{-1/2}$. The data for $N=3200$
indicate that the scaling exponent is slightly closer to the KZS
prediction than the one obtained in a pure linear quench with the same
initial condition and $\tau$-range [cf. Fig. 3(a)]. Since a sudden
quench effectively strengthens the impulse mechanism in the KZS
argument, the number of relevant modes $N_R$ is larger than the one
involved in an adiabatic linear quench. Thus, for the same initial
condition (the same $M_R$), the ratio $\varepsilon$ in
Eq.~(\ref{condition}) is smaller in a sudden quench than in a linear
quench, comparatively leading to a scaling exponent closer to KZS.
Therefore, our conclusions for excited-state sudden quenches are
consistent with the ones reached for excited-state adiabatic quenches,
and reaffirm how small-amplitude sudden quench dynamics and adiabatic
dynamics near a QCP are essentially equivalent over a wide range of
initializations.

\subsection{Adiabatic dynamics following a sudden quench from
the ground state}
\label{ExcSt}

In addition to eigenstates of the initial Hamiltonian, another
physically relevant class of initial preparations is provided by
pure states that are reachable from the many-body ground state via
a sudden parameter quench of {\it arbitrary amplitude}.  For
concreteness, let us focus on adiabatic dynamics following a
sudden quench of the magnetic field $h$ to its critical value
$h_c$ in the Ising chain. Thus, the initial state for the
adiabatic quench is a superposition of different eigenstates of
the Hamiltonian $H(h_c)$ after the (instantaneous) sudden quench.
Since for each mode $k$ the parity quantum number ${\cal P}_k$ is
conserved, and the ground state of $H_k$ lies in the even sector
${\cal P}_k=1$, the expansion coefficients $c_{2,k}=c_{3,k}=0$,
whereas $c_{0,k}$ and $c_{1,k}$ are obtained from expanding the
ground state before the sudden quench in the eigenstate basis $\{
|\psi_k^j (t_0^+)\rangle\}$ of the quenched Hamiltonian $H(h_c)$.

\begin{figure}[t]
\centering
\includegraphics[width=9cm]{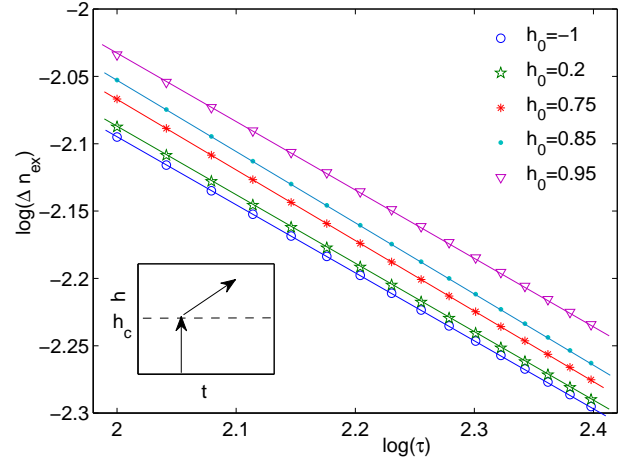}
\caption{(Color online) Scaling of the final relative excitation
density in an adiabatic magnetic-field quench across QCP {\tt A} in
the Ising chain, starting from an excited state prepared by suddenly
quenching $h_0 \mapsto h_c$ for different initial values of $h_0$.
The combined control path is illustrated in the inset. The linear
fitting slope for $h_0=-1, 0.2, 0.75, 0.85, 0.95$ is $-0.50283 \pm
5.0\times10^{-5}$, $-0.50697 \pm 6.0\times10^{-5}$, $-0.5237 \pm
1.0\times10^{-4}$, $-0.52800 \pm 5.0\times10^{-5}$, and $-0.5037 \pm
8.0\times10^{-4}$ respectively. In all cases, the system size
$N=400$.}
\label{fig6}
\end{figure}

We can picture the resulting dynamics in terms of a combined
sudden-adiabatic quench process (see Fig.~\ref{fig6}, inset), except
that unlike in Sec. \ref{Suddenex} we only focus on the scaling
behavior of the relative excitation density $\Delta n_{\text{ex}}(t)$
created after the sudden quench.  Exact numerical results are plotted
in the main panel of Fig.~\ref{fig6}, showing that for a large range
of sudden-quench initializations, the final excitation density still
obeys the same KZS,
$$\Delta n_{\text{ex}} (t_f) \sim \tau^{-d \nu/(\nu z+1)} \sim
\tau^{-1/2},$$
\noindent
as in adiabatic dynamics starting from the ground state.  The above
scaling result can be derived analytically in two limiting cases,
starting from Eq. (\ref{Pexpure}). Upon integrating over all the
relevant modes, we find
\begin{eqnarray}
\Delta n_{\text{ex}}(t) \hspace{-1.5mm}&=&\hspace{-1.5mm}\frac{1}{\pi}
\hspace{-1mm}\int_0^{k_{\text{max}}}\hspace{-5mm} \Delta P_k(t)
dk=\hspace{-1mm}\int_0^{k_\text{{max}}}\hspace{-1mm}
\Big\{(2|c_{0,k}|^2-1)|a_{0,k}(t)|^2 \nonumber \\ &+&
2\text{Re}[c_{0,k}c_{1,k}^\ast a_{0,k}^\ast(t)a_{1,k}(t)]\Big\}
\frac{dk }{\pi}.
\label{eqex}
\end{eqnarray}
There are two contributions in $\Delta P_k(t)$. If the initial state
of mode $k$ is close to either a non-excited or to a fully excited
state ($|c_{0,k}|^2\approx 1$ or $|c_{0,k}|^2\approx 0$ for all $k\in
K_R$, respectively), the first term is the dominant one. In this case,
KZS clearly holds. In the opposite limit where each mode $k\in K_R$ is
initially half-excited ($|c_{0,k}|^2\approx 1/2$), the second term is
the dominant one. Since, for a sudden quench to $h_c$, the latter is
the center of the impulse region (recall Fig. 1, top) and at most half
of the impulse region can be crossed, all the relevant modes can at
most be close to half-excitation, making this second limiting case
directly relevant to the sudden-quench state preparation for suitable
$h_0$.  Assuming that $|c_{0,k}|^2\approx 1/2$ and ignoring relative
phases thus yields
$$\Delta P_k(t) \sim |a_{0,k}(t)a_{1,k}(t)| \sim
|a_{0,k}(t)|\sqrt{1-|a_{0,k}(t)|^2}.$$
\noindent
By invoking the Landau-Zener formula \cite{Victor2}, the asymptotic
($t_f \rightarrow \infty$) excitation probability for modes near $k_c$
scales like $e^{-2\pi k^2\tau}$ when $t_0 \rightarrow
-\infty$. Starting from QCP {\tt A} (the center of the impulse region)
will not, however, affect the exponential behavior
\cite{Vitanov}. Therefore, $|a_{0,k}(t)|^2 \sim e^{-2\pi k^2\tau}$ as
long as $t_f$ is deep in the adiabatic region, and $1-|a_{0,k}(t)|^2
\sim k^2\tau$. Integrating over the relevant modes then gives the
anticipated KZS result:
$$\int_0^{k_{\text{max}}}\hspace{-4mm} dk\,
|a_{0,k}(t)|\sqrt{(1-|a_{0,k}(t)|^2)} \sim
\hspace{-1mm}\int_0^{\tau^{-1/2}}
\hspace*{-5mm}dk \,k\, \tau^{1/2} \sim \tau^{-1/2},$$
\noindent
where we used the fact that $k_{\text{max}} \sim \tau^{-1/2}$
[Eq. (\ref{kmax})] in the upper integration limit.

\begin{figure}[tb]
\centering
\includegraphics[width=9cm]{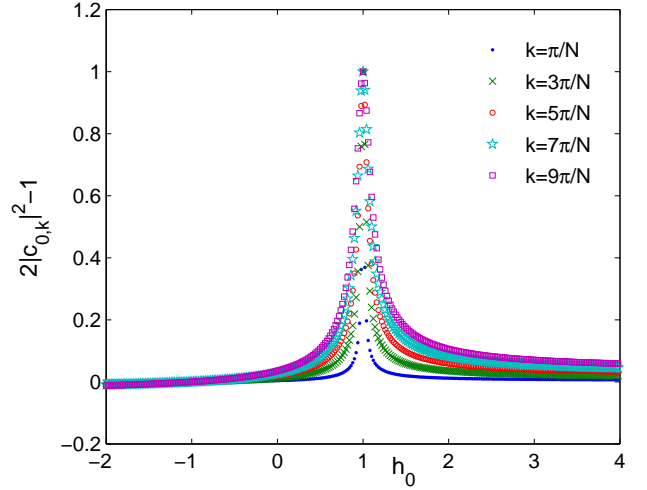}
\caption{(Color online) Dependence of the excitation coefficient
$2|c_{0,k}|^2-1$ upon the initial magnetic-field strength $h_0$ in a
state prepared by a sudden quench $h_0 \mapsto h_c$ in the Ising chain
($\gamma=1$).  The values $c_{0,k}$ are obtained by expanding the ground
state of $H(h_0)$ in terms of the eigenbasis of $H(h_c=1)$ at QCP {\tt A}.
The five lowest-energy modes are considered, for system size $N=400$.}
\label{fig7}
\end{figure}

While the above argument suffices to explain the emergence of KZS
starting from {\em special} sudden-quench initializations, for generic
quenches the dominant term in Eq. (\ref{eqex}) need not be the same
for different modes.  In order to gain further insight, it is
necessary to inspect the distribution of the excitation probability
for each relevant mode after a sudden quench from a generic value $h_0
\mapsto h_c$.  Numerical results for the low-lying modes are presented
in Fig.~\ref{fig7} for a wide range of initial magnetic-field strength
$h_0$.  For each mode $k$, we can identify two boundary values,
$h^{\text{min}}_{0,k}$ and $h^{\text{max}}_{0,k}$, such that when
$h_{0,k}^{\text{min}}\leq h_0 \leq h_{0,k}^{\text{max}}$, mode $k$ is
close to its ground state after the sudden quench ($|c_{0,k}|^2
\approx 1$), whereas if $h_0 \ll h_{0,k}^{\text{min}}$ or $h_0 \gg
h_{0,k}^{\text{max}}$, mode $k$ is close to half-excitation
($|c_{0,k}|^2 \approx 1/2$).  Since $h_{0,k}^{\text{min}}$ and
$h_{0,k}^{\text{max}}$ are approximately symmetric with respect to the
critical value $h_c=1$, let us for simplicity take $h_{0,k}^{\text{m}}
\equiv h_{0,k}^{\text{max}}$, with $h_{0,k}^{\text{min}} \approx 2 h_c
- h_{0,k}^{\text{m}}$.  Qualitatively, $h_{0,k}^{\text{m}}$ can be
determined by the condition $\Delta(h_0,k) \approx \Delta(h_c,k)$,
which yields approximately $|c_{0,k}|^2 \approx 1$.  If, conversely,
$\Delta(h_0,k) \gg \Delta(h_c, k)$ ($|h_0-h_c| \gg
|h_{0,k}^{\text{m}}-h_c|$), we can consider $|c_{0,k}|^2 \approx 1/2$.
Altogether, the results in Figs.~\ref{fig6}-\ref{fig7} indicate that
the limiting analytical condition of requiring the same dominant term
in Eq. (\ref{eqex}) for {\em all} the relevant modes is too strong for
$\Delta n_{\text{ex}}(t)$ to show KZS.  For instance, when $h_0=0.95$,
not all the relevant modes are staying in their ground state ($k_c$ is
not), yet KZS holds.  In general, however, the variation of
$|c_{0,k}|^2$ with $k$ does affect the scaling result. For instance,
when $h_0$ is around $0.75$, agreement with the KZS prediction for the
same system size is relatively poor, motivating one to roughly
identify the range $0.6 \lesssim h_0 \lesssim 0.9$ with a cross-over
region.  Based on these observations, we conjecture that a necessary
(and sufficient) condition for the relative excitation density $\Delta
n_{\text{ex}}(t)$ to approach KZS in the thermodynamic limit is that
{\em the dominant term in Eq. (\ref{eqex}) is the same for the
majority of the relevant modes}.

An alternative physical interpretation of the above conjecture may be
obtained by observing that for a generic value of $h_0$, there exist
modes $k_e, k_g \in K_+$ such that if $k_c \leq k \leq k_e$,
$\Delta(h_0,k) \gg \Delta(h_c,k)$, whilst if $k_g \leq k \leq \pi$,
$\Delta(h_0,k) \approx \Delta(h_c, k)$, and we also assume $k_e < k_g$
for concreteness.  Since, in an adiabatic sweep with speed $\tau$, the
set of relevant modes $K_R=[k_c, k_{\text{max}}]$ is determined
according to Eq. (\ref{kmax}), we can distinguish three different
regimes depending on how $k_{\text{max}}$ is positioned relative to
the interval $[k_e, k_g]$:

(i) $k_{\text{max}} \leq k_e$: in this case, all the relevant modes
are half-excited, recovering one of the limiting situations
(analytically) leading to KZS, as already discussed ({\em e.g.},
$h_0=-1$ in Fig.~\ref{fig6});

(ii) $k_e < k_{\text{max}} < k_g$: in this case, by a reasoning
similar to the one leading to Eq. (\ref{condition}), KZS is predicted
to emerge provided that $(k_{\text{max}}-k_e)=\varepsilon
k_{\text{max}}$, $\varepsilon \ll 1$, in such a way that the majority
of the relevant modes are half-excited ({\em e.g.}, $h_0=0.2$ in
Fig.~\ref{fig6});

(iii) $k_g \leq k_{\text{max}}$: in this case, KZS is predicted to
emerge provided that $k_g=\varepsilon k_{\text{max}}$, $\varepsilon
\ll 1$, in such a way that the majority of the relevant modes stay in
their ground state ({\em e.g.}, $h_0=0.95$ in Fig.~\ref{fig6}).

Thus, for both $h_0=0.75$ and $h_0=0.85$, the initial state prepared
by the sudden quench may be interpreted to lie in the cross-over
region between cases (ii) and (iii), explaining why the resulting
scaling deviates appreciably from the KZ prediction.

Similarly to the excited-eigenstate initialization, sudden-quench
initialization will also add more constraints on the appropriate
$\tau$ range for KZS to hold. If the initial state is prepared via a
sudden quench that guarantees one of the above conditions (i)--(iii)
to be fulfilled for any $\tau \in
[\tau_{\text{min}},\tau_{\text{max}}]$, then the latter range is also
appropriate for KZS to emerge under excited-eigenstate initialization.
If not, the situation is more involved, and the range of $\tau$ may
need to be adjusted such that either (ii) or (iii) is enforced.  If
condition (ii) is more likely to be obeyed ({\em e.g.}, if $h_0
\approx 0.6$), we can choose $\tilde{\tau}_{\text{min}}>
\tau_{\text{min}}$ in such a way that the number of modes between
$k_e$ and $k_g$ is decreased, and the majority of relevant modes is
thus half-excited. If instead condition (iii) is more likely to be
obeyed ({\em e.g.}, if $h_0 \approx 0.9$), we can choose $
\tilde{\tau}_{\text{max}} < \tau_{\text{max}}$ in such a way that the
number of relevant modes staying in their ground state is
increased. While the strategy for adjusting the $\tau$-range in a
sudden-quench initialization is similar to the one advocated in
excited-eigenstate initialization, conditions (i)--(iii) are in fact
easier to fulfill than Eq.~(\ref{condition}). For instance, for
$N=400$, the worst scaling in Fig.~\ref{fig6} is still relatively
close to KZS, whereas the latter is completely lost when initially
only $k_c$ is excited in Fig.~\ref{fig3}(a). This difference is due to
the fact that the initial occupation of modes in the relevant set
changes less abruptly in a sudden-quench initialization than in
excited-eigenstate initialization.

We conclude our discussion of quench processes originating from a
(pure) excited state by commenting on the fact that the analysis
developed for $\Delta n_{\text{ex}}(t)$ can be extended to different
observables without requiring major conceptual modifications.  While
an explicit example involving the spin correlator defined in
Eq. (\ref{dxx}) will be included in the next Section, the basic idea
is to proceed in analogy with ground-state quenches \cite{Deng2}, by
taking into consideration the appropriate scaling exponent as
determined by the physical dimension of the observable ${\cal O}$.
Consider, for instance, the relative excitation energy $\Delta H(t)$
defined in Eq. (\ref{dH}) which, as remarked, can be experimentally
more accessible than the relative excitation density.  In all the
situations where KZS holds for the latter, $\Delta n_{\text{ex}}(t_f)
\sim \tau^{-d\nu/(\nu z +1)} \sim \tau^{-1/2}$ (in particular, in the
case of excited-state initialization via a sudden quench just
discussed), we also find for our model that
$$ \Delta H(t_f) \sim \tau^{-(d+z)\nu/(\nu z +1)} \sim \tau^{-1},$$
\noindent
consistent with the corresponding ground-state scaling behavior
explored in Refs.~\onlinecite{Deng2,Deng3}.

\section{Quenches from a thermal state}
\label{Thermal}

\subsection{Adiabatic quench dynamics}
\label{thermaladiab}

While we have only focused thus far on initialization mechanisms
resulting in a {\em pure} excited state, another large class of
initial states with a finite excitation energy may be obtained through
dissipative means, in particular because the system may find itself
(or be placed) in contact with a thermal bath. After a time sufficient
for equilibration to occur, the system would then relax to a canonical
ensemble at temperature $T$. In equilibrium, it is well known that the
influence of a ground-state QCP can cross over to a finite range of
temperatures, the so-called ``quantum critical regime,'' which is
often broader than naively expected \cite{Sachdev1,Sachdev,Gegenwart}.
In a dynamical scenario, how robust is dynamical scaling (in
particular, KZS) to initialization at a finite-temperature?  If
scaling persists, how do the relevant non-equibrium exponents depend
upon the initial temperature?  Motivated by these questions, scaling
behavior in a system initially prepared in a thermal equilibrium state
{\em at criticality} and then adiabatically quenched away from the QCP
has been analyzed in Ref.~\onlinecite{DeGrandi}.  In particular, it is
shown that for fermionic quasi-particles, the excess excitation due to
a quench across a standard QCP obeys
\begin{equation}
\Delta n_{\text{ex}} (t_f) \sim \frac{1}{T}
\,\tau^{-{(d+z)\nu}/(\nu z+1)} ,
\label{therm1}
\end{equation}
provided that the initial temperature is high enough ($T \gg
\epsilon_k(t_0)$, for all $k \in K_R$).  Our goal here is to both
present quantitative evidence for the above scaling law and, most
importantly, to extend the analysis to multicritical QCPs.

Let $T$ denote the initial thermal equilibrium temperature, so that
the initial density operator has the form $\rho(t_0)=\bigotimes_{k \in
K_+} \rho_k(t_0)$, with $\rho_k(t_0)$ given by:
\begin{eqnarray}
\rho_{00,k}(t_0)& \hspace*{-1mm}=\hspace*{-1mm}& \frac{1}{{\cal Z}}
e^{+\epsilon_k(h,\gamma)/T}, \;\; \rho_{11,k}(t_0) =\frac{1}{{\cal Z}}
e^{-\epsilon_k(h,\gamma)/T}, \nonumber \\ \rho_{22,k}(t_0)&
\hspace*{-1mm}= \hspace*{-1mm}& \rho_{33,k}(t_0)=\hspace{1mm}
\frac{1}{{\cal Z}},
\label{thermstate}
\end{eqnarray}
in units where $\hbar=k_B=1$ and with
$${\cal Z}\equiv
2+e^{+\epsilon_k(h,\gamma)/T}+e^{-\epsilon_k(h,\gamma)/T}.$$
\noindent
For clarity, we focus on linear adiabatic dynamics first. We shall
study both the standard Ising QCP {\tt A} under a magnetic-field
quench of the form $h(t)=1-t/\tau$ [$h=h_c=1, \gamma=1$ in
Eq. (\ref{thermstate})], and the MCP {\tt B} under a simultaneous
quench of the magnetic field and the anisotropic parameter,
$h(t)=1-\gamma(t)=1-t/\tau$ [$h=h_c=1, \gamma=\gamma_c=1$ in
Eq. (\ref{thermstate})].  At $T=0$, the scaling of the excitation
density can be in both cases described by $n_{\text{ex}}(t_f) \sim
\tau^{-{d \nu z}/[z_2(\nu z+1)}]$, where $z_2$ is determined from the
scaling of the minimal gap along the path with respect to $k$
[cf. Eq. (4) in Ref. \onlinecite{Deng3}, with $\alpha=1$ and
$d_2=0$]. Thus, $z_2=z$ in the quench across QCP {\tt A}, leading to
KZS, whereas $z_2=3\neq z$ in the quench across MCP {\tt B}, leading
to anomalous scaling $n_{\text{ex}} (t_f) \sim \tau^{-1/6}$.  Given
the above thermal initial condition, starting from
Eq. (\ref{Pexmixed}) for the relative excitation probability, one
finds:
\begin{eqnarray}
\Delta P_k
(t)=\tanh\Big(\frac{\epsilon_k(h_c,\gamma_c)}{2T}\Big)|a_{0,k}(t)|^2,
\end{eqnarray}
\noindent where for both paths we simply write
$\epsilon_k(h_c,\gamma_c)$ to mean that critical parameter values
are assumed at $t=t_0$. When $T \leq \epsilon_k(h_c,\gamma_c)$,
$\tanh\big(\frac{\epsilon_k(h_c,\gamma_c)}{2T}\big) \approx 1$ and
$\Delta P_k (t)$ is the same as starting from the ground state of
mode $k$. Thus, in order for the same ground-state scaling (either
KZS or $\tau^{-1/6}$) to emerge in the low-temperature limit, the
condition $T \leq \epsilon_k(h_c,\gamma_c)$ needs to be satisfied
for all the relevant modes.  Since
$\epsilon_{k_c}(h_c,\gamma_c)=0$, this means that in the
thermodynamic limit, the only allowed initial temperature is $T=0$
if a thermal state of $H(h_c,\gamma_c)$ is considered.  In the
opposite limit of high temperature, where $T \gg
\epsilon_k(h_c,\gamma_c)$,
$\tanh\big(\frac{\epsilon_k(h_c,\gamma_c)}{2T}\big) \approx
\epsilon_k(h_c,\gamma_c)/(2T) \sim (k-k_c)^z/T$ for modes $k$ near
$k_c$. Upon integrating over the relevant modes and recalling Eq.
(\ref{kmax}), the relative excitation density is then:
\begin{eqnarray}
\Delta n_{\text{ex}}(t_f)&=&\frac{1}{\pi} \int_0^{k_{\text{max}}}
\Delta P_k (t_f) d^d k \sim \frac{1}{T} \int_0^{\tau^{-\nu z /[z_2
(\nu z+1)]}}
\hspace*{-15mm}k^z d^dk \nonumber \\
&=&
\frac{1}{T}\,\tau^{-(d+z)\nu z /[z_2 (\nu z+1)]}.
\label{therm1MCP}
\end{eqnarray}
For the standard QCP {\tt A}, this yields $\Delta n_{\text{ex}} (t_f)
\sim \tau^{-1}/T$, recovering the result of Eq. (\ref{therm1}), whilst
$\Delta n_{\text{ex}} (t_f) \sim \tau^{-1/2}/T$ in the multicritical
quench across QCP {\tt B}.  In Ref.~\onlinecite{Deng3}, we argued that
the time-dependent excitation process in ground-state quenches need
not be dominated by the critical mode $k_c$ for certain paths across
MCPs and $P_k=\Delta P_k \sim k^{d_2}$, with $d_2$ playing the role of
an ``effective dimensionality exponent''.  For a thermal quench, it is
interesting to note that, formally, one may interpret $d_2=z\neq 0$ in
the above equation, also implying that the dominant contribution does
{\em not} originate from modes around $k_c$. In the high temperature
limit, $\rho_{k}(t_0)$ is, indeed, almost fully mixed for modes near
$k_c$, causing the contribution of $\rho_{00,k}$ and $\rho_{11,k}$ to
Eq.  (\ref{Pexmixed}) to be nearly cancelled, and consistently leading
to $\Delta P_{k}(t) \approx 0$ for those modes.

\begin{figure}[t]
\centering
\includegraphics[width=9cm]{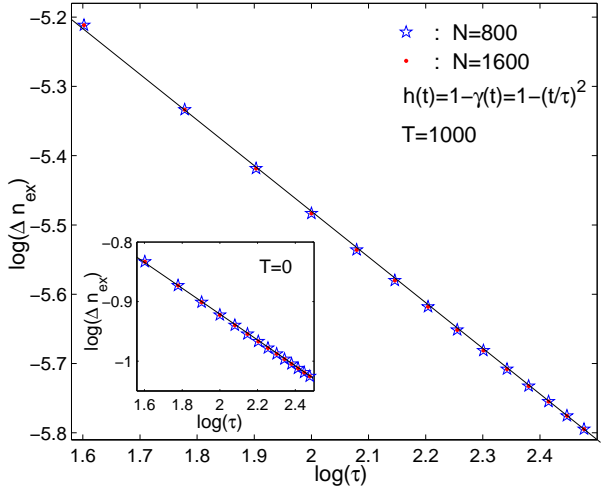}
\caption{(Color online) Exact scaling behavior of $\Delta
n_{\text{ex}}(t_f)$ in a quadratic adiabatic quench
$h(t)=1-\gamma(t)=1-(t/\tau)^2$, starting from a thermal state at MCP
{\tt B} ($t_0=t_c=0$) toward the FM phase. The initial temperature
$T=1000$, yielding a linear fitting slope $-0.663 \pm 0.002$, in good
agreement with the value $2/3$ predicted by Eq. (\ref{therm2}).  For
comparison, the case of a ground-state quench is reproduced in the
inset, with a linear fitting slope of $-0.2190 \pm 0.0006$, which is
also in good agreement with the predicted $2/9$ exponent \cite{Deng3}.
The data for different sizes ($N=800$ and $N=1600$) coincide up to
$10^{-13}$.}
\label{fig8}
\end{figure}

\begin{figure}[t]
\centering
\includegraphics[width=9cm]{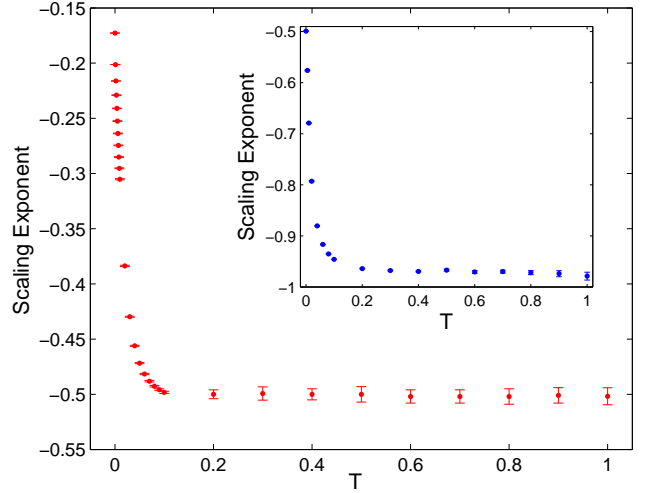}
\caption{(Color online) Main panel: Scaling exponent $\Delta XX(t_f)$
as a function of temperature $T$ in a linear quench
$h(t)=1-\gamma(t)=1-t/\tau$ away from the MCP {\tt B}, starting with a
thermal equilibrium state of $H(h_c,\gamma_c)$. Inset: Scaling
exponent of $\Delta XX(t_f)$ as a function of temperature in a linear
quench $h(t)=1+t/\tau$ away from the regular QCP {\tt A}, starting
with a thermal equilibrium state of $H(h_c=1)$. In both cases, the
system size $N=800$. }
\label{fig9}
\end{figure}

The scaling prediction in Eq. (\ref{therm1MCP}) can be further
generalized to a {\em non-linear thermal quench}, whereby for instance
$h(t)=1-\gamma(t)=1-(t/\tau)^\alpha$ in the case of a quench away from
the MCP {\tt B}.  When $T=0$, Eq. (4) in Ref. \onlinecite{Deng3}
yields \cite{RemarkAmit} $n_{\text{ex}}(t_f) \sim \tau^{-{d \alpha \nu
z}/{[z_2(\alpha\nu z+1)]}}.$ Correspondingly, in the high-temperature
limit,
\begin{equation}
\Delta n_{{\text{ex}}}(t_f) \sim \frac{1}{T} \,
\tau^{-{(d+z)\alpha \nu z}/[{z_2(\alpha\nu z+1)]}}.
\label{therm2}
\end{equation}
\noindent
Exact numerical results for a quadratic quench ($\alpha=2$) are
reported in Fig.~\ref{fig8}, the inset corresponding to the
ground-state $T=0$ case.  Within numerical accuracy, the observed
behavior is in excellent agreement with the predicted scaling,
$\tau^{-2/9}$ for $T=0$ and $\tau^{-2/3}$ for high-$T$, respectively.

We further examine how dynamical scaling is detected by other
observables and how it is influenced by temperature away from the
limiting regimes discussed above by considering the behavior of the
spin correlator, $\Delta XX(t)$, defined in Eq. (\ref{dxx}).  Since
$XX$ does not commute with the Hamiltonian in Eq. (\ref{Ham}), no
analytical treatment is possible. Exact numerical results are
presented in Fig.~\ref{fig9} for both the regular and the
multicritical QCPs {\tt A} and {\tt B} (inset vs. main panel,
respectively), starting from the same thermal initial condition at
criticality as considered above.  As the data show, similar features
emerge in both cases: the scaling exponent of $\Delta XX(t)$, which is
expected to be the same as for $\Delta n_{{\text{ex}}}$, deviates from
its zero-temperature value (${-1/2}$ or ${-1/6}$, respectively) as
soon as the temperature is nonzero, and as the latter is gradually
increased, it continuously changes until for sufficiently high
temperature ($T \gg \epsilon_k(h_c,\gamma_c)$, for all $k \in K_R$),
it stabilizes at the value predicted by Eq. (\ref{therm2}) (${-1}$ or
${-1/2}$, respectively).  All these observations are consistent with
the predictions in the previous paragraph.

In summary, ground-state dynamical scaling (and KZS in particular) is
fragile with respect to temperature fluctuations if the initial state
is a thermal equilibrium state {\em at criticality}.  In this case,
the two situations where scaling exists are the zero-temperature and
the high-temperature limit, with Eq. (\ref{therm2}) holding in the
latter regime.  This requires {\em all} the relevant modes to either
stay in their ground state or be highly mixed at the initial time,
which is a stronger condition in comparison to the ones identified in
the previous sections for coherently-prepared (pure) excited states.
From a practical standpoint, the high-temperature regime could
potentially be relevant to liquid-state NMR simulators \cite{Qsim}.
In order for tests of dynamical scaling/KZS in the low-temperature
regime to be experimentally viable, however, the initial thermal state
needs to be (or be prepared) sufficiently {\em far away from
criticality} ({\em e.g.}, $|h_0-h_c| \gg 1$ for QCP {\tt A}), in such
a way that the condition $T \leq \Delta(k,h_0)$ for all $k \in K_R$
can still be fulfilled with a non-zero temperature.

\subsection{Sudden quench dynamics and thermalization}
\label{effectivetherm}

Sudden quenches have recently attracted considerable interest as a
setting for probing the long-time dynamics of isolated many-body
systems and the approach to equilibrium
\cite{Cardy,Rossini,Sei,Rossini2,Paolo,Canovi}.  Since the quadratic
Hamiltonian in Eq. (\ref{Ham}) describes a simple (non-interacting)
integrable model, it is well known that no thermalization can occur in
a proper sense, that is, the behavior of {\em generic} observables is
not governed by a conventional statistical equilibrium ensemble
\cite{thermalization,Rigol}.  The above investigations have
nevertheless shown that information about the asymptotic behavior of
an appropriate subset of observables may still be encoded in a finite
{\em effective temperature} $T_{\text{eff}}$, independent on the fine
details of the initial state and the dynamics but only determined by
the total energy of the process.  Let $\rho(t_0)\equiv \rho_0$ and
$H_f$ denote, respectively, the density operator describing the
initial state of the system, and the final Hamiltonian after the
(instantaneous) quench.  Following Rossini {\em et
al.}~\cite{Rossini}, the effective temperature is defined by the
requirement that the average energy of the initial state relative to
the quenched Hamiltonian equals the one corresponding to a {\em
fictitious} thermal state at temperature $T_{\text{eff}}$ in the
canonical ensemble, that is,
\begin{eqnarray}
\text{Tr} [\rho_0 H_f ]=\text{Tr} [\rho_{T_{\text{eff}}} H_f].
\label{teff}
\end{eqnarray}
Under the assumption that $T=0$ initially [that is, ground-state
initialization in Eq. (\ref{teff})], the emergence of effective
thermal behavior has been related to the {\em locality} properties of
different physical observables relative to the quasi-particle language
that diagonalizes the model~\cite{Sei,Rossini2}.  For a generic quench
in a Ising chain, only non-local observables (such as the two-point
correlation functions of the order parameter) have been found to
thermalize, with both their asymptotic average value and the
finite-time transient being determined by {\em equilibrium}
statistical mechanics at $T_{\text{eff}}$.  Remarkably, however,
thermal behavior has also been established for certain {\em local}
observables (the transverse magnetization per site, $1/N\sum_j
\sigma_z^j$, and the kink density, ${\cal N}$) {\em in quenches
towards criticality}, the long-time value being still univocally
determined by $T_{\text{eff}}$.

Physically, it is clear that the concept of an effective temperature
has a restricted validity and, for the model under investigation, it
does not imply that an actual thermal ensemble emerges as a result of
a sudden quench followed by free evolution under the quenched
Hamiltonian.  
With that in mind, we further explore in what follows the emergence of
effective thermal behavior in {\em critical} quenches, by focusing on
a different local observable and by extending the analysis in two
directions: first, initialization in a thermal state at finite $T>0$
and, second, sudden quenches to a multicritical QCP.

\begin{figure}[t]
\centering
\includegraphics[width=9.2cm]{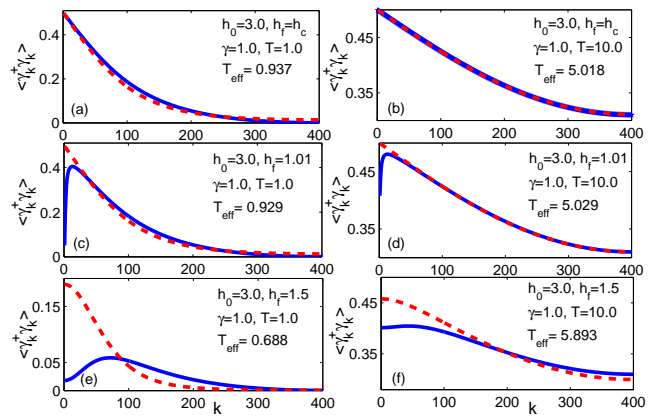}
\caption{(Color online) Comparison between the long-time average
quasiparticle excitation following a sudden quench $h_0\mapsto h_f$
starting from a thermal initial state at temperature $T$ (dashed red)
and the equilibrium value predicted by a fictitious thermal canonical
ensemble at $T_{\text{eff}}$ (solid blue). Panels (a), (c), (e):
sudden quenches to $h_f=h_c=1$, $h_f=1.01$, $h_f=1.5$, respectively,
with initial temperature $T=1.0$. The behavior for a ground-state
quench ($T=0$, data not shown) is qualitatively similar, with
deviations from the thermal prediction being further pronounced.
Panels (b), (d), (f): sudden quenches to $h_f=h_c=1$, $h_f=1.01$,
$h_f=1.5$, respectively, with initial temperature $T=10.0$. In all
cases, $N=800$, and the value of $T_{\text{eff}}$ obtained from is
Eq. (\ref{teff}) is also given. }
\label{fig10}
\end{figure}

Let us first consider a sudden quench of the magnetic field $h_0
\mapsto h_f$ in the Ising chain ($\gamma=1$), starting from an initial
state of the form given in Eq. (\ref{thermstate}), and focus on the
long-time behavior of the number of quasiparticle excitations with
momentum $k$.  Since the corresponding observable commutes with the
time-dependent Hamiltonian, the long-time expectation value
$\langle\gamma_k^\dag \gamma_k\rangle$ coincides with the one right
after the quench. In order for the latter to be consistent with the
equilibrium value at $T_{\text{eff}}$, the following identity must
hold:
\begin{eqnarray}
(\rho_{00,k}(t_0)&-&\rho_{11,k}(t_0))|a_{0,k}|^2+\rho_{11,k}(t_0)
+\rho_{33,k}(t_0) \nonumber \\ &=& (1+
e^{+\epsilon_k(h_f,\gamma=1)/T_{\text{eff}}})^{-1},
\label{teffex}
\end{eqnarray}
where $|a_{0,k}|^2$ is the excitation probability of mode $k$ due to
the quench and Eq.~(\ref{rhot}) has been used in the left
hand-side. The right hand-side is the fermionic thermal equilibrium
prediction Tr$[\rho^k_{T_{\text{eff}}} \gamma_k^\dag \gamma_k]$. Exact
numerical results are presented in Fig.~\ref{fig10}. Altogether, these
data indicate that similar to the behavior of other local observables
in a ground-state quench~\cite{Sei,Rossini2}, no effective
thermalization is observed outside criticality [panels (c)--(f)], as
expected. Even for a quench toward QCP {\tt A}, however, the initial
temperature $T$ {\em must be sufficiently high} in order for our
chosen observable to thermalize [panel (a) vs.  (b)].

\begin{figure}[t]
\centering
\includegraphics[width=8cm]{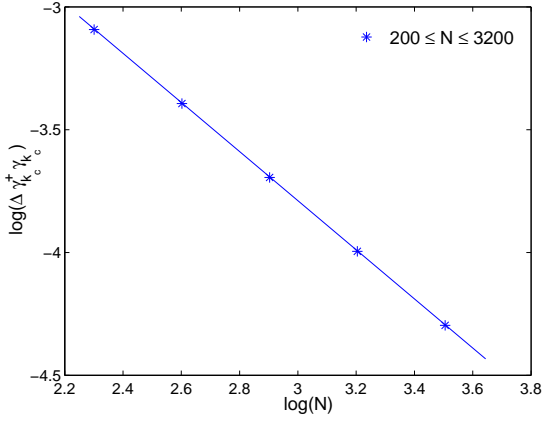}
\caption{(Color online) Difference between the long-time quasiparticle
excitation of the critical mode $k_c$ from its thermal equilibrium
prediction as a function of system size for a sudden magnetic-field
quench to $h_c$ in the Ising chain. An initial thermal state with
temperature $T=1.0$ is considered. The linear fitting slope is
$-0.99992 \pm 3\times 10^{-5}$.}
\label{fig11}
\end{figure}

In order to gain physical insight into what distinguishes a critical
vs. non-critical quench in our case, and understand why effective
thermal behavior fails to emerge outside criticality even for high
initial temperature, it is useful to take a closer look at
Fig.~\ref{fig10}(d): clearly, the main difference between the
equilibrium and the actual quasiparticle distribution arises from
momentum modes close to $k_c$.  On the one hand, since
$\Delta(k_c,h_f)$ is the smallest gap at $h_f$, the maximum
quasiparticle excitation is expected to occur at $k_c$ from the
equilibrium prediction [right hand side of Eq. (\ref{teffex})]. On the
other hand, the peak of the observed distribution is located at modes
close to $k_c$, but not exactly at $k_c$.  Because the system is far
from $h_c$, note that the difference of $\rho_{00,k}$, $\rho_{11,k}$,
and $\rho_{33,k}$ for modes close to $k_c$ is negligible.  Thus, the
main difference is due to $|a_{0,k}|^2$, which, as remarked, is the
excitation probability of mode $k$ at $T=0$ after a sudden quench to
$h_c$.  Upon re-interpreting $|a_{0,k}|^2 \leftrightarrow
1-|c_{0,k}|^2$, it is possible to make contact with the results shown
in Fig.~\ref{fig7}: clearly, the excitation probability of mode $k_c$
changes dramatically for $h_0$ close to $h_c$, which suggests that
$k_c$ does not contribute appreciably {\em unless} $h_f=h_c$.
Instead, other modes close to $k_c$ can be excited for values
$h_f\approx h_c$ at which $k_c$ is not yet excited.  Since the
excitation contribution from such ``quasi-critical modes'' would then
be larger than the one from $k_c$, Eq.~(\ref{teffex}) would not
hold. Accordingly, the only way to enforce the validity of
Eq.~(\ref{teffex}) is through a sudden quench towards $h_c$, as
observed.

Having clarified why criticality is essential, we turn to assess
whether the requirement of a sufficiently high initial $T$ may be
related to the finite system size or will persist in the thermodynamic
limit.  We focus on a sudden quench $h_0=3.0 \mapsto h_c$ at $T=1.0$,
and analyze how the long-time average of the total quasiparticle
density $1/N \sum_k \gamma_k^\dag \gamma_k$ deviates from the thermal
equilibrium prediction at $T_{\text{eff}}$ as $N$ is increased.  While
we find that the observed deviations are practically constant over the
range of $N$ explored (data not shown), the difference between
$\langle \gamma_{k_c}^\dag \gamma_{k_c}\rangle$ and its corresponding
thermal prediction at $T_{\text{eff}}$ does decrease with increasing
$N$: as seen in Fig.~\ref{fig11}, such a difference $\Delta
\gamma_{k_c}^\dag \gamma_{k_c} \sim N^{-0.99992}$ at $T=1.0$, implying
a vanishing difference and effective thermal behavior also at low
temperature {\em for the critical mode} as $N \rightarrow \infty$.
This property, however, stems from the fact that the gap of $k_c$
closes in the thermodynamic limit, which is not true for the gap of
other modes.  For either the number of quasiparticles in a generic
mode or for the total quasiparticle density, we thus conjecture that
even in the thermodynamic limit, thermal behavior will be observed
following sudden quenches to the QCP {\tt A} {\em only if $T \gg
\Delta_k(h_c,\gamma=1)$ for all the relevant modes}.

\begin{figure}[t]
\centering
\includegraphics[width=9cm]{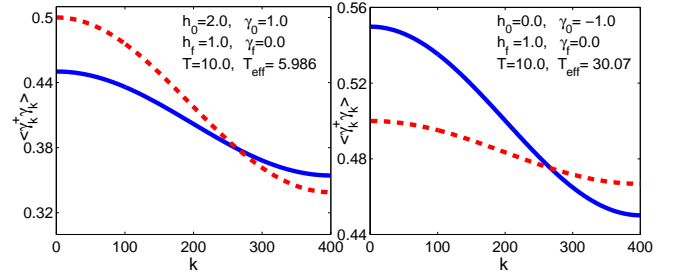}
\caption{(Color online) Comparison between the long-time quasiparticle
excitation following a sudden quench $h_0\mapsto
h_f=h_c=1,\gamma_0\mapsto \gamma_f=\gamma_c=0$ towards the MCP {\tt B}
(blue) and the equilibrium value predicted by a fictitious thermal
ensemble at $T_{\text{eff}}$ (red).  The system is initially in a
thermal state with temperature $T=10$. Left: Initial state is the
thermal state at $h_0=2.0, \gamma_0=1.0$ (inside the PM phase).
Right: Initial state is the thermal state at $h_0=0.0,\gamma_0=-1.0$
(inside the FM phase).  Notice that due to the fact that the
excitation probability of low-energy modes exceeds $1/2$,
$T_{\text{eff}}$ is much higher than in any other situation with the
same initial $T$, cf. Fig.~\ref{fig10}(b, d, f) and
Fig.~\ref{fig12}(a).  }
\label{fig12}
\end{figure}

In view of the peculiar features that distinguish a multicritical QCP,
as reflected in particular in anomalous scaling behavior \cite{Deng3},
it is not obvious whether the above condition would still suffice for
the same observables to thermalize in a sudden quench towards MCP {\tt
B}. Exact results for two sudden multicritical quenches of the form
$(h_0=1+\gamma_0 \mapsto h_f =1+\gamma_f, \gamma_0\mapsto \gamma_f)$
are given in Fig.~\ref{fig12}, starting from a thermal state at high
temperature: specifically, MCP {\tt B} is both reached via a sudden
quench from the PM phase (left panel) and via a sudden quench from the
FM phase (right panel).  Contrary to the high-temperature scenario for
the regular QCP {\tt A} [Fig. \ref{fig10}(b)], {\em no} thermal
behavior emerges, the observed expectation value $\langle
\gamma_k^\dag \gamma_k \rangle$ for modes close to $k_c$ being
significantly smaller or larger than the thermal equilibrium
prediction, respectively.

This anomalous long-time behavior can be traced back to the asymmetry
of the impulse region along the control path, as sketched in
Fig.~\ref{impulse}(bottom).  Following Ref.~\onlinecite{Deng3}, the
location of the minimum gap for each mode $k$ along the path
$h=1+\gamma$ is determined by requiring
$\partial{\Delta_k(\gamma,1+\gamma)}/\partial{\gamma}=0$, that is,
$$\tilde{\gamma}(k)=(\cos{k}-1)/(1+\sin{k}^2) < 0,$$ which indicates
that the center of the impulse region is largely shifted into the FM
phase for each $k$. As a result, after a sudden quench to the MCP {\tt
B} from the FM phase, the excitation probability of low-energy modes
tends to be enhanced above $1/2$, whereas for a sudden quench to MCP
{\tt B} from the PM phase, the excitation probability of low-energy
modes tends to be suppressed below $1/2$.  Since the thermal
equilibrium value is close to $1/2$ in the high-temperature limit for
low-energy modes, thermal behavior is not realized in either quench
process.

Based on the above results, we conjecture that {\em quenching toward
the center of the impulse region} is a necessary requirement for
$\gamma_k^\dag \gamma_k$ or $1/N \sum_k \gamma_k^\dag \gamma_k$ to
thermalize following a sudden quench.  While typically this is the
case in a quench to a regular QCP (for instance, a sudden quench of
$h$ to QCP {\tt A} at fixed $\gamma=1$), for a sudden quench to MCP
{\tt B} along the path $h=1+\gamma$, the location of the minimum gap
(hence the center of the impulse region) is different for each mode
$k$, preventing thermalization to be possible along this path
irrespective of the final values $h_f,\gamma_f$.  More generally, we
expect the above requirement to be {\em necessary} for local
observables other than those examined here.  While this goes beyond
our current scope, it would be interesting to verify, for instance,
whether the transverse magnetization or the density of kinks would
still effectively thermalize in a multicritical quench to QCP {\tt B}
from the ground state.

We also remark that in a recent work \cite{Paolo}, general conclusions
have been reached for the equilibrium distribution after a sudden
quench, predicting, in particular, effective thermal behavior for
generic observables when the quench is performed around a non-critical
point, and poor equilibration otherwise.  While at first these results
seem to contradict both our present conclusions for critical quenches
towards QCP {\tt A} in the appropriate temperature regime as well as
earlier results for zero temperature \cite{Sei,Rossini2}, a crucial
assumption in Ref. \onlinecite{Paolo} is a small quench amplitude,
causing only a small number of excited states to effectively
contribute around a QCP.  The opposite condition is implied throughout
our discussion, the sudden quench amplitude being in fact large enough
for the number of excited states involved in a critical quench to
outweigh those involved in a non-critical one (cf. Fig.~\ref{fig7}).
In the light of that, we also conjecture that having a sufficiently
large number of states involved in the excitation process is a general
necessary condition for effective thermalization after a sudden
quench.

\section{Conclusions}
\label{conclusion}

In summary, we have addressed how different aspects of many-body
non-equilibrium dynamics depend upon initialization in a state other
than the ground state for a class of one-dimensional exactly solvable
XY models.  Our main findings may be itemized as follows:

{\bf $\bullet$ Dynamical scaling: initial pure excited states.}
Provided that the non-equilibrium response of the system is
characterized in terms of suitable {\em relative} indicators (such as
the relative excitation density), adiabatic quench dynamics can still
encode the ground-state equilibrium critical exponents for a large
class of initial energy eigenstates as well as for pure excited states
prepared by a sudden parameter quench.  A crucial role is played by
how the initial excitation is distributed over the set of {\em
relevant} quasi-particle modes that effectively evolve in an adiabatic
quench.  In particular, a unifying criterion that ensures the
emergence of KZS in both the above scenarios in the thermodynamic
limit is obtained by requiring that the {\em majority of the relevant
modes share a common initial excitation pattern}, as expressed by
Eq. (\ref{condition}).

Our results recover ground-state scaling when no excitation is
initially present, but they also allow for the critical exponents of
the ground-state QPT to be encoded in the scaling behavior for highly
energetic initial configurations, where most of the relevant modes are
fully or half-excited.  While this makes contact with similar
conclusions on critical entanglement scaling in excited energy
eigenstates recently obtained for {\em static} QPTs \cite{Alba}, it
confirms that only the set of relevant modes is important in dynamical
scenarios, as opposed to the full static mode set.  Beside being
supported by exact numerical methods and analytical derivations in
limiting regimes, a justification of the proposed criterion has also
been obtained for the case of excited-eigenstate initialization by
suitably extending the perturbative (first-order) AR approach we
previously employed for ground-state continuous QPTs.

{\bf $\bullet$ Dynamical scaling: initial mixed states.}  In general,
more restrictive conditions on the distribution of the initial
excitation over relevant modes must be obeyed for universal dynamical
scaling to emerge in adiabatic quench dynamics that originates from a
statistical (incoherent) mixture as compared to a (coherently
prepared) pure state.  In particular, two distinct scaling regimes
have been identified for an initial thermal ensemble at a finite
temperature $T$, depending on whether the latter is very low or very
high relative to the relevant quasiparticle energy scale, and leading
to KZS $\tau^{-1/2}$ vs. $\tau^{-1}$ for a standard QCP,
respectively. Since in both cases {\em all} the relevant modes must
share a common excitation pattern if the initial thermal state is
prepared {\em at criticality}, this implies that KZS is fragile
against thermal fluctuations in this case, the scaling exponent
deviating from the KZ prediction as soon as $T\ne 0$.  From a
practical standpoint, it is however important to note that a finite
range of temperatures can still support KZS if the system is initially
at thermal equilibrium sufficiently far from criticality.  General
predictions for scaling behavior in adiabatic thermal quenches
involving a MCP have also been obtained [cf. Eq. (\ref{therm2})], and
verified to be consistent with exact numerical results.

{\bf $\bullet$ Effective thermalization.}  Effective thermal behavior
may emerge in the relaxation dynamics of the quasiparticle density
following a sudden quench from a thermal state under appropriate
conditions.  Specifically, the long-time expectation value of this
observable is determined by a {\em fictitious} thermal equilibrium
ensemble at temperature $T_{\text{eff}}$ provided that i) the system
is quenched toward the {\em center of the impulse region}, and ii) the
initial temperature is {\em sufficiently high} with respect to all the
relevant gaps.  For a standard QCP, the first requirement is met by a
sudden quench {\em toward criticality}, which has been found
sufficient for local observables such as the transverse magnetization
per site and the kink density to thermalize starting from the ground
state \cite{Rossini2,Sei}.  Our results indicate that, in general,
condition i) alone need {\em not} suffice for arbitrary local
observables.  While requirement ii) may be taken to be in line with
what expected for a free (integrable) theory \cite{Cardy,Rigol}, it
remains an interesting open question to precisely characterize what
subclass of local observables may ehxibit effective thermal behavior
under the sole condition i) \cite{localobs}.

Our results additionally show that for certain observables (such as
the quasiparticle density), effective thermalization may fail to occur
altogether (or possibly require yet more stringent requirements) for
sudden quenches to a {\em multicritical} QCP.  Physically, we have
traced this behavior back to the existence of quasicritical
(path-dependent) energy states and the corresponding shift of the
impulse region, which also underlies the emergence of anomalous
scaling exponents \cite{Deng3}. In this context, an interesting next
step would be to examine the thermalization behavior of other local
observables as considered in Ref. \onlinecite{Rossini2,Sei}.

While the above analysis provides a more complete picture of
non-equilibrium dynamics in a paradigmatic class of spin chains than
available thus far, it remains a main open question to understand how
crucially our results rely on the XY chain being an exactly solvable
non-interacting model. From this point of view, it would be worthwhile
to explore, for instance, whether dynamical critical scaling may still
exist for finite-energy initial states in non-integrable models, or
even in more complex but still integrable systems such as a
Bethe-Ansatz solvable one-dimensional Heisenberg XXZ chain \cite{Alba}
or an infinitely coordinated Lipkin-Meshkov-Glick model \cite{Caneva}.

\vspace*{5mm} 

\acknowledgments

S.D. gratefully acknowledges partial support from a Hull Graduate
Fellowship.


{}


\begin{thebibliography}{100}
\bibitem{RemarkAmit} We note that our prediction differs from the one
recently established in Ref.~\onlinecite{Amit}, where a linearization
procedure around the QCP is invoked in order to use the Landau-Zener
formula.
\bibitem{localobs} From this point of view, it is interesting to note
that both the transverse magnetization per site and the density of
kinks are closely related to the final quenched Hamiltonian [cf. in
particular Eq. (35) in Ref.  \onlinecite{Rossini2}].
\end{thebibliography}
\end{document}